\documentstyle[mnextra]{mn}

\def\KDelta{\Delta_{\rm Kron}}
\def\CalkDelta{\Delta_{\rm cal}^{\rm Kron}}
\def\CaleDelta{\Delta_{\rm cal}^{\rm extrap.}}
\def\ConvDelta{\Delta_{\rm conv}}
\def\Mpc{{\rm Mpc}}
\def\Lsun{{\rm L_\odot}}
\def\Msun{{\rm M_\odot}}
\def\gsim{\ga}
\def\eg{e.g.\ }

\def\etal{{et al.\ }}
\def\alphaj{\alpha_{\rm J}} 
\def\alphak{\alpha_{\rm K_S}} 
\def\Mstarj{{M^\star_{\rm J}}} 
\def\Mstark{{M^\star_{\rm K_S}}} 
\def\Mstar{{M^\star}} 
\def\phij{{\Phi^\star_{\rm J}}} 
\def\phik{{\Phi^\star_{\rm K_S}}} 
\def\twomass{{2MASS}}
\def\twodF{{2dFGRS}} 
\def\K{K$_{\rm S}$} 
\def\B{\rm b$_{\rm J}$} 
\def\nmtot{40,121}     
\def\nztot{17,173}     

\def\nztottwodF{140,000}  %
\def\datetwodF{September 2000} %
\def\fracone{88.6}        
\def\fractwo{11.4}        
\def\fracthree{6.7}   
\def\fracfour{9.6}       

\def\fracsix{9.3}      
\def\fracseven{90.7}    
\def\fracredshift{42.8} 

\def\ifracstars{49.5}   
\def\ifracmrg{47.6}     
\def\ifracmisc{2.9}    
\def\fracstars{4.6} 
\def\fracmrg{4.4}     
\def\fracmisc{0.27}     

\def\areatwodF{1094.727   +     740.3251  + 3.16557*100}

\def\areatwodF{2151.6}
\def\areaeff{619}      
\def\areaeffa{642}      
      
\input{epsf}
\nobrackets    
\seceqnum      
\begin{document}

\title[Near IR Luminosity Functions]{The 2dF Galaxy Redshift Survey:
  Near Infrared Galaxy Luminosity Functions\thanks{This publication makes
  use of data products from the Two Micron All Sky Survey (2MASS),
  which is a joint project of the University of Massachusetts and the
  Infrared Processing and Analysis Center/California Institute of
  Technology, funded by the National Aeronautics and Space
  Administration and the National Science Foundation.}
\vspace*{-0.5 truecm} 
}

\author[S. Cole et al.] {
\parbox[t]{\textwidth}{Shaun Cole$^1$, Peder Norberg$^1$, 
Carlton M. Baugh$^1$, Carlos S. Frenk$^1$, Joss Bland-Hawthorn$^2$, 
Terry Bridges$^2$,
Russell Cannon$^2$, Matthew Colless$^3$, Chris Collins$^4$, 
Warrick Couch$^5$, Nicholas Cross$^6$,
Gavin Dalton$^7$, Roberto De Propris$^5$,
Simon P. Driver$^6$, 
George Efstathiou$^8$, Richard S. Ellis$^9$, Karl Glazebrook$^{10}$, 
Carole Jackson$^3$,
Ofer Lahav$^8$, Ian Lewis$^2$, Stuart Lumsden$^{11}$, 
Steve Maddox$^{12}$, Darren Madgwick$^8$,
John A. Peacock$^{13}$, Bruce A. Peterson$^3$,
Will Sutherland$^{13}$, Keith Taylor$^2$ (The 2dFGRS Team)}
\vspace*{6pt} \\ 
$^1$Department of Physics, University of Durham, Science Laboratories, South 
Road, Durham DH1 3LE, United Kingdom \\
$^2$Anglo-Australian Observatory, P.O. Box 296, Epping, NSW 2121, Australia \\
$^3$Research School of Astronomy \& Astrophysics, The Australian National 
University, Weston Creek, ACT 2611, Australia \\
$^4$Astrophysics Research Institute, Liverpool John Moores University, Twelve 
Quays House, Egerton Wharf, Birkenhead, L14 1LD, UK \\
$^5$Department of Astrophysics, University of New South Wales, Sydney, 
NSW2052, Australia \\
$^6$School of Physics and Astronomy, North Haugh, St Andrews, Fife, KY16 9SS,
United Kingdom \\
$^7$Department of Physics, Keble Road, Oxford OX1 3RH, United Kingdom \\
$^8$Institute of Astronomy, University of Cambridge, Madingley Road, 
Cambridge CB3 0HA, United Kingdom \\
$^9$Department of Astronomy, California Institute of Technology, Pasadena, 
CA 91125, USA \\
$^{10}$Department of Physics \& Astronomy, Johns Hopkins University, 3400 
North Charles Street Baltimore, MD 21218\-2686, USA \\
$^{11}$Department of Physics \& Astronomy, E C Stoner Building, Leeds LS2 9JT,
United Kingdom \\
$^{12}$School of Physics and Astronomy, University of Nottingham, University 
Park, Nottingham, NG7 2RD, United Kingdom \\
$^{13}$Institute of Astronomy, University of Edinburgh, Royal Observatory, 
Edinburgh EH9 3HJ, United Kingdom \\
\vspace*{-1.0 truecm}
}

\maketitle

\begin{abstract}
We combine the \twomass\ extended source catalogue and the 2dF galaxy
redshift survey to produce an infrared-selected galaxy catalogue with
\protect{\nztot} measured redshifts. We use this extensive dataset 
to estimate the galaxy luminosity functions in the J- and \K-bands.  The
luminosity functions are fairly well fit by Schechter
functions with parameters $\Mstarj-5 \log h= -22.36 \pm 0.02$,
$\alphaj= -0.93\pm 0.04$, $\phij=0.0104 \pm 0.0016 h^3$ Mpc$^{-3}$ in
the J-band and $\Mstark- 5 \log h= -23.44 \pm 0.03$, $\alphak=
-0.96\pm 0.05$, $\phik=0.0108 \pm 0.0016 h^3$ Mpc$^{-3}$ in the
\K-band (\twomass\ Kron magnitudes).  These parameters are derived assuming a
cosmological model with $\Omega_0=0.3$ and $\Lambda_0=0.7$.  
With datasets of this size, systematic rather than random errors
are the dominant source of uncertainty in the determination of the 
luminosity function. We carry out a careful investigation of possible
systematic effects in our data.
The surface brightness distribution of the sample shows
no evidence that significant numbers of low surface brightness or
compact galaxies are missed by the survey.  We estimate the
present-day distributions of \B$-$\K\ and J$-$\K\ colours as a function of
absolute magnitude and use models of the galaxy stellar populations,
constrained by the observed optical and infrared colours, to infer the
galaxy stellar mass function.  Integrated over all galaxy masses, this
yields a total mass fraction in stars (in units of the critical mass
density) of $\Omega_{\rm stars}h= (1.6 \pm 0.24) \times 10^{-3}$ for a Kennicutt
IMF and $\Omega_{\rm stars}h= (2.9 \pm 0.43)\times 10^{-3}$ 
for a Salpeter IMF. These values are consistent with those inferred from
observational estimates of the total star formation history of the universe
provided that dust extinction corrections are modest.
\end{abstract}

\begin{keywords}
galaxies: luminosity function estimators 
\vspace*{0.5 truecm}
\hrule
\vspace*{-1.5 truecm}
\end{keywords}

\section{Introduction}

The near-infrared galaxy luminosity function is an important
characteristic of the local galaxy population. It is a much
better tracer of evolved stars, and hence of the total stellar content
of galaxies, than optical luminosity functions which can be 
dominated by young stellar populations and are also strongly affected
by dust extinction. Hence, infrared luminosities can be much more
directly related to the underlying stellar mass of galaxies and so
knowledge of the present form and evolution of the 
infrared galaxy luminosity function 
places strong constraints on the history of star formation in the 
universe and on galaxy formation models (\eg Cole \etal
\shortcite{cole2000} and references therein). 

The local K-band luminosity function has been estimated from optically 
selected samples by Mobasher, Sharples and Ellis (\shortcite{mobasher93}),
Szokoly \etal (\shortcite{szk}) and Loveday (\shortcite{love}) and from K-band
surveys by Glazebrook \etal (\shortcite{kgb95}), and Gardner \etal 
(\shortcite{gard97}). The existing K-band surveys are small. 
The largest, by Gardner et al., covers only 4~deg$^2$ and contains only
510 galaxies. 
The recent survey of Loveday covers a much larger solid angle.
In this survey the redshifts were known in advance of measuring
the K-band magnitudes and this was exploited by targetting bright
and faint galaxies resulting in an effective sample size 
much larger than the 345 galaxies actually measured. However,
like all optically selected samples, it suffers from
the potential problem that galaxies with extremely red infrared to
optical colours could be missed. In this paper we combine the
2-Micron All Sky Survey (\twomass) with the 2dF galaxy redshift 
survey (\twodF) to create an infrared selected redshift survey 
subtending \areatwodF~deg$^2$.
Currently the sky coverage of both surveys is incomplete, but already the
overlap has an effective area of \areaeff~deg$^2$. Within this area
the redshift survey is complete to the magnitude limit of the \twomass\
catalogue and so constitutes a complete survey which is
50~times larger than the previous largest published infrared 
selected redshift survey. A new catalogue of a similarly large area,
also based on \twomass, has very recently been analysed by
Kochanek \etal (\shortcite{kochanek2001}). They adopt isophotal
rather than total magnitudes and concentrate on the 
dependence of the luminosity function on galaxy morphology.

In Section~\ref{sec:selection} 
we briefly describe the relevant properties of the
\twodF\ and \twomass\ catalogues. Section~\ref{sec:matching} is a detailed
examination of the degree to which the matched
\twomass--\twodF\ galaxies are a complete and representative subset of 
the \twomass\ catalogue.
Section~\ref{sec:mags} examines the calibration of the 
\twomass\ total magnitudes and Section~\ref{sec:compl}
demonstrates that the \twomass\ catalogue and the inferred luminosity
functions are not affected by surface brightness selection effects.
In Section~\ref{sec:pop} we present the method by which we compute
k-corrections and evolutionary corrections and relate the observed
luminosities to the underlying stellar mass. The estimation methods
and normalization of the luminosity functions are described briefly in
Section~\ref{sec:lf_est}. Our main results are presented and discussed in
Section~\ref{sec:results}.  These include estimates of the J and \K\
(K-short) luminosity functions, the \B$-$\K\ and J$-$\K\ colour
distributions as a function of absolute magnitude and the distribution
of spectral type. We also estimate
the stellar mass function of galaxies, which can be integrated to infer
the fraction of baryons in the universe which are in
the form of stars.  We conclude in Section~\ref{sec:conc}.

\section{The Dataset}
\label{sec:data}

The data that we analyze are the extended source catalogue from the
second incremental release of the 2-Micron All Sky Survey (\twomass\
http://pegasus.phast.umass.edu) and the galaxy catalogue of the 2dF
galaxy redshift survey (\twodF\ http://www.mso.anu.edu.au/2dFGRS).
Here, we present the relevant properties of these two catalogues and
investigate their selection characteristics and level of completeness.

\subsection{Selection Criteria}
\label{sec:selection}

The \twomass\ is a ground-based, all-sky imaging survey in the 
J, H and \K\ bands. Details of how extended sources are identified
and their photometric properties measured are given by
Jarrett \etal (\shortcite{jarrett00}). The detection sensitivity (10$\sigma$)
for extended sources is quoted as 14.7, 13.9 and 13.1 magnitudes in 
J, H and \K\ respectively. The complete survey is expected to contain
1~million galaxies of which approximately 580,000 are contained
in the second incremental data release made public in March 2000.

The \twodF\ is selected in the photographic \B\ band
from the APM galaxy survey
(\cite{apmI},\shortcite{apmII},\shortcite{apmIII}) and subsequent
extensions to it, that include a region in the northern galactic cap
(Maddox \etal in preparation). The
survey covers approximately \areatwodF~deg$^2$ consisting of  two broad
declination strips. The larger is centred on the SGP and approximately
covers $-22^\circ\negthinspace.5$$>$$\delta$$>$$-37^\circ\negthinspace.5$,
$21^{\rm h}40^{\rm m}$$<$$\alpha$$<$$3^{\rm h}30^{\rm m}$; 
the smaller strip is in
the northern galactic cap and covers
$2^\circ\negthinspace.5$$>$$\delta$$>$$-7^\circ\negthinspace.5$, 
$9^{\rm h}50^{\rm
m}$$<$$\alpha$$<$$14^{\rm h}50^{\rm m}$.  In addition, there are a number of
randomly located circular 2-degree fields scattered across the full extent
of the low extinction regions of the southern APM galaxy survey.  There are
some gaps in the \twodF\ sky coverage within these  boundaries due to
small regions that have been excluded around bright stars and satellite
trails. The \twodF\ aims to measure the redshifts of all the  galaxies
within these boundaries with extinction-corrected \B\ magnitudes brighter than
19.45.  When complete, at the end of 2001, 250,000 galaxy redshifts will
have been measured.  In this paper we use the \nztottwodF\ redshifts 
obtained prior to \datetwodF .

The overlap of the two surveys is very good.  There are some gaps in
the sky coverage due to strips of the sky that were not included in
the \twomass\ second incremental release, but overall a substantial
fraction of the \areatwodF~deg$^2$ of the \twodF\ is covered by
\twomass .  The homogeneity and extensive sky coverage of the combined
dataset make it ideal for studies of the statistical properties of the
galaxy population.

\begin{figure}
\centering
\centerline{\epsfxsize=8 truecm \epsfbox[0 270 550 770]{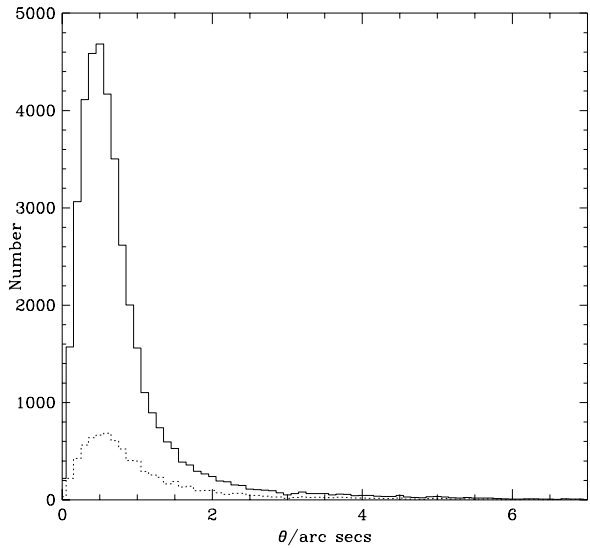}}
\caption{The distribution of angular separation, $\theta$, for
matched \twomass--\twodF\ galaxies. The solid  histogram is the
distribution for the whole catalogue and the dotted histogram for
the subset of \twomass\ galaxies with semi-major axes larger
than 12~arcsec.
}
\label{fig:seps}
\end{figure}

\subsection{The Completeness of the Matched \twomass--\twodF\ Catalogue}
\label{sec:matching}

Here we consider whether all the \twomass\ galaxies within the \twodF\
survey region have \twodF\ counterparts and assess the extent to which the
fraction of galaxies with measured redshifts  represents
an unbiased sub-sample.

The astrometry in both \twomass\ and \twodF\ is, in general, very good 
and it is an easy matter to match objects in the two catalogues.
We choose to find the closest pairs within a search radius equal to
three quarters of the semi-major axis of the J-band image 
(denoted j\_r\_e in the \twomass\ database). Scaling the search 
radius in this way helps with the matching of large extended objects.
This procedure results in the identification of \twodF\ counterparts 
for \nmtot\ of the \twomass\ objects, when at random one would only expect to 
find a handful of such close pairs. Moreover, the distribution of separations
shown in Fig.~\ref{fig:seps} peaks at 0.5~arcsec, with only 3\% having 
separations greater than 3~arcsec. A significant part of this tail
comes from the most extended objects as is evident from
the dotted histogram in Fig.~\ref{fig:seps} which shows objects
with semi-major axes larger than 12~arcsec.
Thus, we can be very confident in these identifications.

\begin{figure*}
\centering
\centerline{\epsfxsize=15.0 truecm \epsfbox[20 50 530 750]{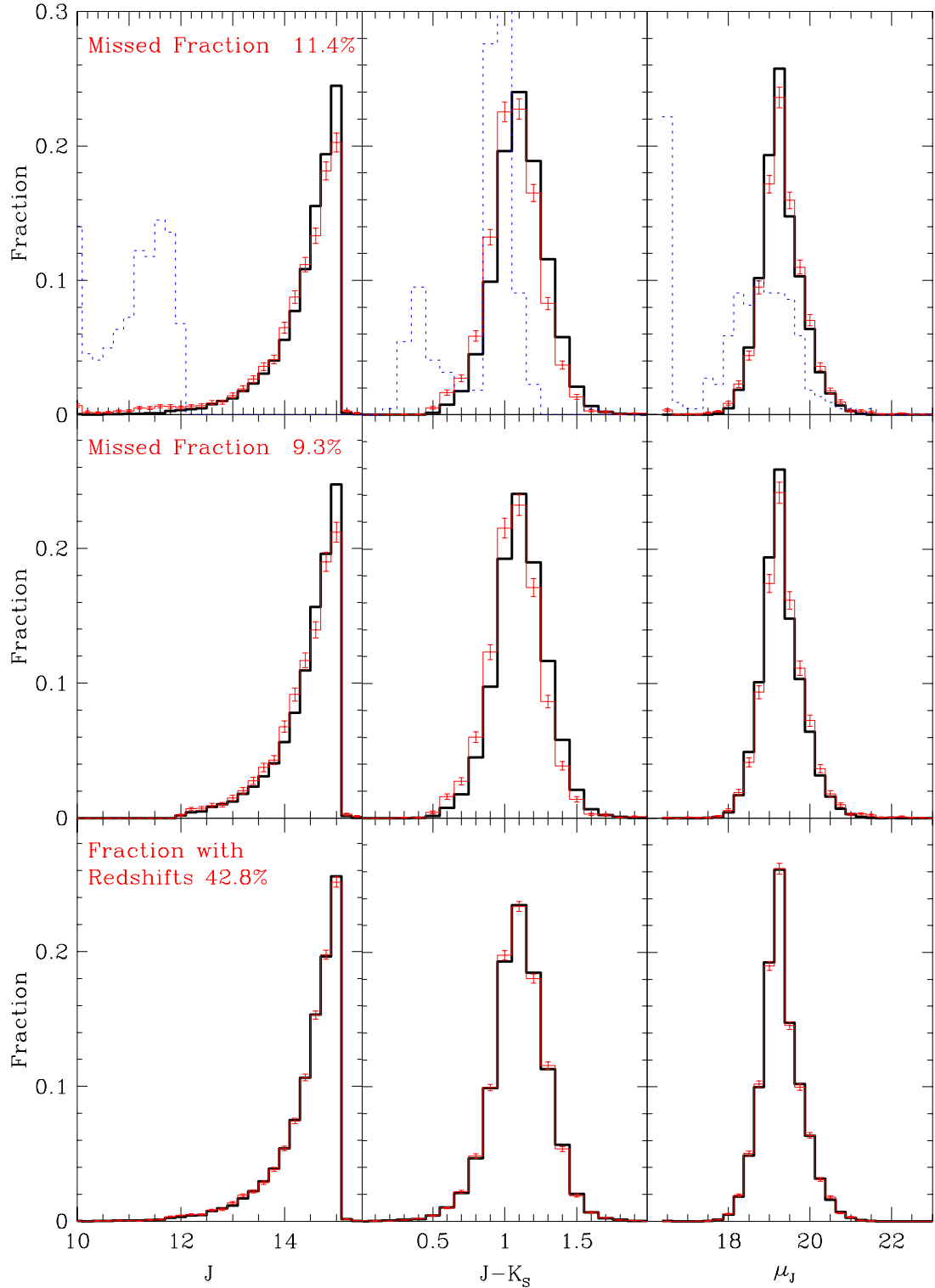}}
\caption{The distribution of J-band apparent magnitude, J$-$\K\ colour and
J-band surface brightness, $\mu_{\rm J}$, 
for various sub-samples of the \twomass\ catalogue. Here, the measure
of surface brightness used is simply 
$\mu_{\rm J}\equiv {\rm J} - 5 \log_{10}r$, 
where J is the Kron magnitude and $r$ the Kron semi-major axis in arc seconds
(j\_m\_e and j\_r\_e in the 2MASS database). 
In all three rows, the thick solid histograms are the distributions for
\twomass\ objects that are matched with \twodF\ galaxies. The light solid
histograms in the top row are the \fractwo \% of \twomass\ galaxies
that are not matched with \twodF\ galaxies. Poisson errorbars are shown on 
these histograms.  The dashed histograms are for the bright sub-sample with
J$<$12. In the middle row, the light histograms show the distributions
for the \fracsix \% of \twomass\ galaxies fainter than J$=$12
and satisfying the additional image classification constraints discussed
in the text that are not matched with \twodF\ galaxies. 
In the bottom row, the light histograms show the distributions
of the \fracredshift \% of the matched \twomass--\twodF\ galaxies 
for which redshifts have been measured. The values in each histogram are
the fraction of the corresponding sample that falls in each bin.
}
\label{fig:hists}
\end{figure*}

The \nmtot\ \twomass\ objects for which we have found secure
\twodF\ counterparts amount to \fracone \% of the \twomass\ extended
sources that fall within the boundary of the \twodF.  As discussed
below, a more restrictive criterion that includes only sources fainter
than J$=$$12$ that are confidently classified as galaxies by \twomass , 
increases the fraction with \twodF\ matches to
\fracseven \%. The remaining \fracsix \% are missed for well
understood reasons (star-galaxy classification:
\fracstars \%; merged or close images: \fracmrg \%; miscellaneous:
\fracmisc \%), none of which ought to introduce a bias. This is
confirmed explicitly, in the middle row of Fig.~\ref{fig:hists}, by the
close correspondence between the photometric properties of the missed
\fracsix \% and those of the larger matched sample.  Hence, in
estimating luminosity functions no significant bias will be introduced
by assuming the matched sample to be representative of the full
population. Furthermore, the distribution shown in the bottom row of
Fig.~\ref{fig:hists} shows that the subset of \protect{\nztot} galaxies
for which we have measured redshifts is a random sample
of the full matched \twomass--\twodF\ catalogue.
This summary is the result of a thorough investigation, which
we describe in the remainder of this section, into the reasons 
why \fractwo \% of the \twomass\ sources are missed and whether their
omission introduces a bias in the properties of the matched sample.  

We first consider objects in the \twomass\ catalogue which based on
their images and colours are not confidently classified as galaxies.
In the \twomass\ database a high e\_score or g\_score indicates a high
probability that the object is either not an extended source or not a
galaxy. A cc\_flag$\ne$0 indicates an artifact or contaminated and/or
confused source. For detailed definitions of these parameters we refer
the reader to Jarrett \etal (\shortcite{jarrett00}). Rejecting all 
objects which have either e\_score$>$1.4, g\_score$>$1.4 or
cc\_flag$\ne$0 removes just \fracthree \% of the total.  However,
removing these reduces significantly the fraction of the \twomass\
sample that does not match with the \twodF\ catalogue, from \fractwo
\% to \fracfour \%. Thus, it is likely that about 30\% of the
\twomass\ objects which have e\_score$>$1.4, g\_score$>$1.4 or
cc\_flag$\ne$0 are not galaxies.

\begin{figure}
\centering
\centerline{\epsfxsize=8.5 truecm  \epsfbox[0 410 530 750]{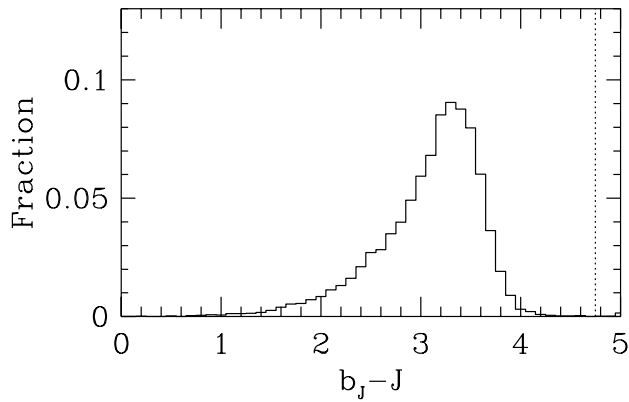}}
\caption{The solid histogram shows the 
distribution of \B-J colours for \twomass\ galaxies selected
to have J$<$$14.7$. (Here, we use the 
\twomass\ default magnitude, denoted j\_m in the \twomass\
database.) The vertical dashed line indicates the colour at which
this sample starts to become incomplete due to the \B$<$$19.45$ magnitude
limit of \twodF. 
}
\label{fig:colours}
\end{figure}

The \twomass\ may contain a tail of very red objects that are too faint
in the \B-band to be included in the \B$<$19.45 \twodF\
sample. Fig.~\ref{fig:colours} shows the distribution of \B-J colours
for the matched objects with J$<$14.7. (Here, the J-band magnitude we
are using is the default magnitude denoted j\_m in the \twomass\
database. In Section~\ref{sec:mags} we will consider the issue of what
magnitude definition is most appropriate for estimating the
luminosity function.)  The vertical dashed line indicates the colour
at which this sample starts to become incomplete due to the \B$<$19.45
magnitude limit of \twodF. The colour distribution cuts off sharply
well before this limit, suggesting that any tail of missed very red
objects is extremely small.  In other words the \twodF\ is
sufficiently deep that even the reddest objects detected at the
faintest limits of \twomass\ ought to be detected in \twodF.

In the top row of Fig.~\ref{fig:hists} we compare the 
distributions of magnitude, colour and surface brightness
for the matched and missed \twomass\ objects. In general, the properties
of the missed subset overlap well with those of the much larger
matched subset. However, we do see that the distributions for missed
objects contain tails of bright and  blue objects. It is quite
likely that this is due to the \twomass\ extended source catalogue being 
contaminated by a  small population of saturated  or multiple stars. 
The dotted histograms in the top row of
Fig.~\ref{fig:hists} show the distributions of magnitude, colour 
and surface brightness for the bright subset of the missed objects with
J$<$12. Here we clearly see bimodal colour and surface brightness
distributions. The blue peak of the colour distribution is consistent
with that expected for stars (see Jarrett \etal \shortcite{jarrett00}).
Excluding these bright, J$<$12, objects which are clearly contaminated
by stars reduces the fraction of missed \twomass\ objects from
\fracfour \% to \fracsix \%. The magnitude, colour and surface brightness
distributions for this remaining \fracsix \% are shown in the middle
panel of Fig.~\ref{fig:hists}. We see that the missed objects are
slightly under-represented at the faintest magnitudes and also
slightly bluer on average than the matched sample, while the distribution 
of surface brightness is almost indistinguishable for the two 
sets of objects. These differences are small and so will introduce
no significant bias in our luminosity function estimates. 

To elucidate the reasons for the remaining missed \fracsix \% of
\twomass\ objects we downloaded 100 1$\times$1~arcmin  images from
the STScI Digitized Sky Survey (DSS) centred on the positions of a
random sample of the missed \twomass\ objects. In each image we
plotted a symbol to indicate the position of any \twodF\ galaxies
within the 1$\times$1~arcmin field.  We also plotted symbols to
indicate the positions and classifications of all images identified in
the APM scans from which the \twodF\ catalogue was drawn, down
to a magnitude limit of \B$\approx 20.5$.  These images are classified
as galaxies, stars, merged images (galaxy+galaxy, galaxy+star or star+star) or
noise.  This set of plots allows us to perform a census of the reasons
why some \twomass\ objects are not present in the \twodF\ survey.

The main cause for the absence of \twomass\
objects in the \twodF\ is that the APM has classified
these objects as stars. These amount to \ifracstars \% of the missed sample
(\fracstars \% of the full \twomass\ sample).  In some cases, the DSS
image shows clearly that these are stars and in others that they are
galaxies. However, the majority of these objects cannot easily be
classified from the DSS images. Thus, they could be galaxies that the APM
has falsely classified as stars or stars that \twomass\ has falsely
classified as galaxies. The first possibility is not unexpected since
the parameters used in the APM star-galaxy separation algorithm
were chosen as a
compromise between high completeness and low contamination such that
the expected completeness is around 95\% with 5\% stellar
contamination (\cite{apmI}). It is hard to rule out the
possibility that this class of object does not include a substantial
fraction of stars, but if so, their presence appears not to distort
the distribution of colours shown in Fig~\ref{fig:colours}. Another
\ifracmrg \% of the random sample (\fracmrg \% of the full \twomass\ sample)
are classified by the APM as mergers or else consist of two close
images in the DSS but are classified by the APM as a single galaxy offset from
the \twomass\ position.  The remaining \ifracmisc \% of the random sample
(\fracmisc \% of the full \twomass\ sample) are missed for a variety
of reasons including proximity to the diffraction spikes of very
bright stars and poor astrometry caused by the presence of a neighbouring
unclassified image.

\subsection{\twomass\ Magnitude Definitions and Calibration}
\label{sec:mags}

\begin{figure}
\centering
\centerline{\epsfxsize=9.0 truecm \epsfbox[137 27 550 770]{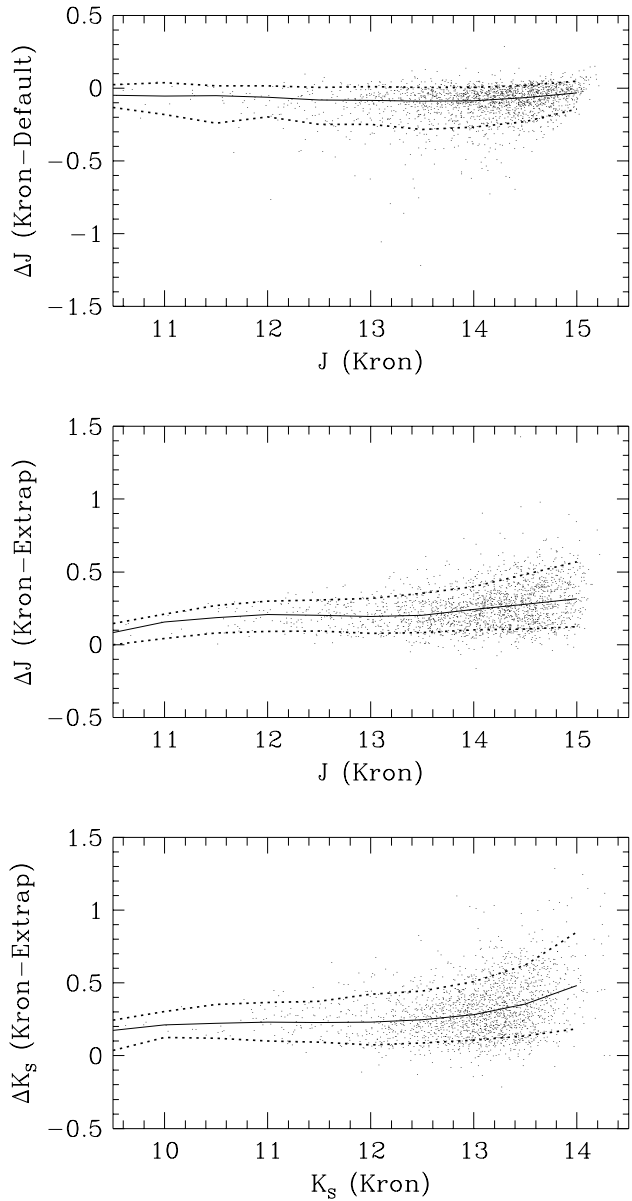}}
\caption{A comparison of the \twomass\ default, Kron and extrapolated
magnitudes in the J and \K\ bands. The dots are the measured values for
each of the galaxies in the matched \twomass--\twodF\ catalogue. The
solid and dotted lines indicate the median, 10 and 90 percentiles of the
distribution.
}
\label{fig:mags}
\end{figure}

\begin{figure*}
\centering
\centerline{\epsfxsize=18.0 truecm \epsfbox[0 255 525 720]{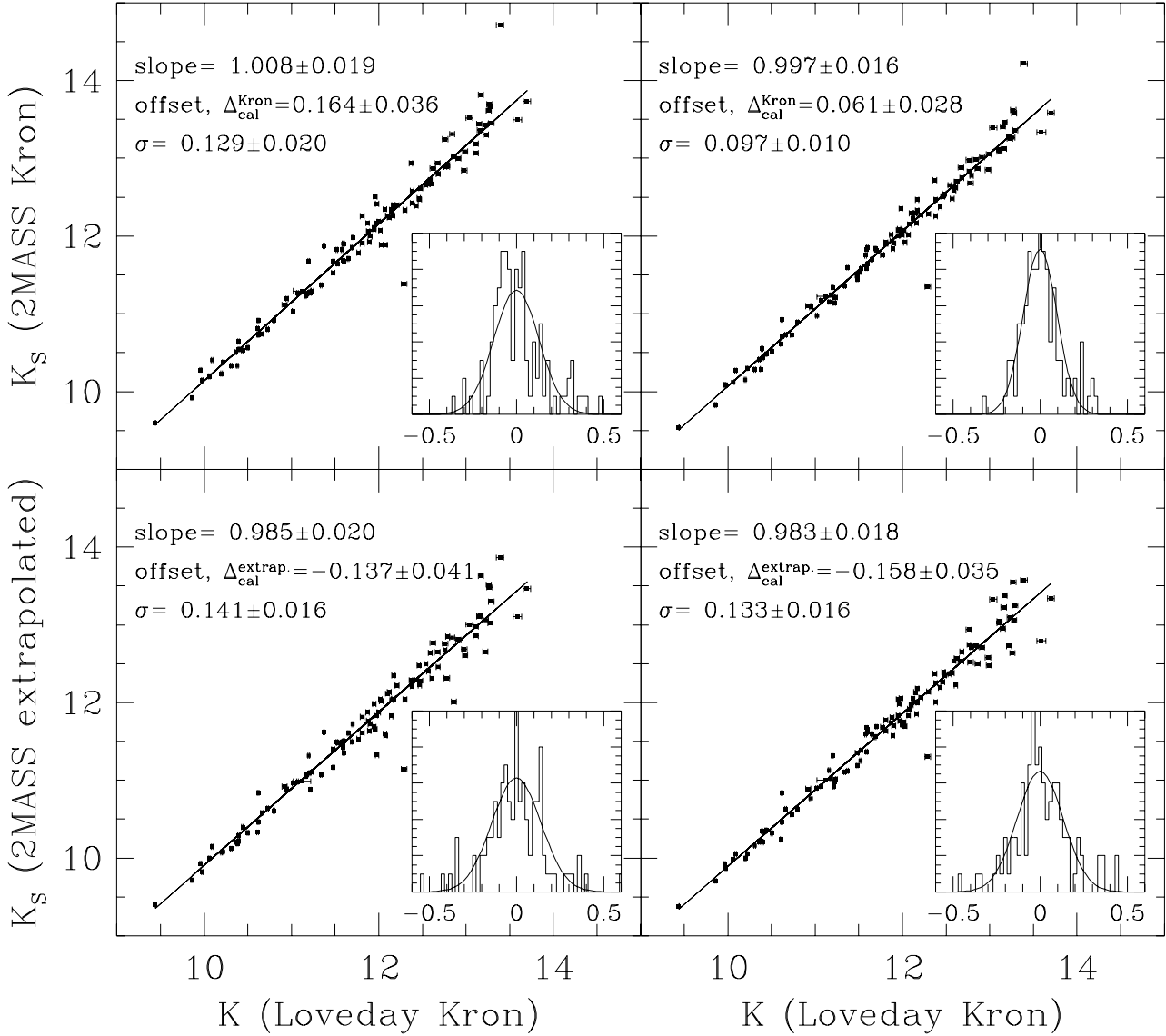}}
\caption{Comparison of \twomass\ Kron and extrapolated magnitudes
with the independent measurements of Loveday (2000). The left 
hand panels are for the \K-band Kron and extrapolated magnitudes
(k\_m\_e and k\_m\_ext in the \twomass\ database).
The right hand panels show Kron
and extrapolated magnitudes inferred from the \twomass\ J-band
Kron and extrapolated magnitudes and the measured default aperture 
J$-$\K\  colours
(j\_m\_e$-$j\_m$+$k\_m and j\_m\_ext$-$j\_m$+$k\_m 
in the \twomass\ database variables). 
The horizontal
errorbars show the measurement errors quoted by Loveday (2000).
The solid lines show simple least squares fits.
The slopes and zero-point offsets of these fits and the rms residuals 
about the fits are indicated on each panel. The inset plots show the
distribution of residual magnitude differences.
}
\label{fig:loveday}
\end{figure*}

The \twomass\ extended source database provides a large selection of
different magnitude measurements. In the previous section we used the
default magnitudes (denoted j\_m and k\_m in the \twomass\
database). These are magnitudes defined within the same circular
aperture in each waveband. For galaxies brighter than \K$=$$14$,  the aperture
is the circular \K -band isophote of 20~mag arcsec$^{-2}$ 
and for galaxies fainter than \K$=$14 it
is the circular J-band isophote of 21~mag arcsec$^{-2}$.
These are not the most useful definitions of magnitude for
determining the galaxy luminosity function. Since we are interested in
measuring the total luminosity and ultimately the total stellar mass
of each galaxy, we require a magnitude definition that better
represents the total flux emitted by each galaxy.  We consider Kron
magnitudes (Kron \shortcite{kron}) and extrapolated magnitudes. Kron
magnitudes (denoted j\_m\_e and k\_m\_e in the \twomass\ database) are
measured within an aperture, the Kron radius, defined as 2.5 times the
intensity-weighted radius of the image. The extrapolated magnitudes
(denoted j\_m\_ext and k\_m\_ext in the \twomass\ database) are
defined by first fitting a modified exponential profile, $f(r) = f_0
\exp [-(\alpha r)^{1/\beta}]$, to the image from 10~arcsec to the
20~mag/arcsec$^2$ isophotal radius, and extrapolating this from the
Kron radius to 4 times this radius or 80~arcsec if this is smaller
(Jarrett private communication). Note that improvements are being made
to the extended source photometry algorithms developed and employed
\twomass\ team and so in the final \twomass\ data release the definitions 
of the Kron and extrapolated magnitudes may be slightly modified
(Jarrett private communication).

Fig.~\ref{fig:mags} compares the default, Kron and extrapolated 
magnitudes in the J and \K\ bands for the matched \twomass--\twodF\ catalogue.
The upper panel shows that while the median offset between the J-band isophotal
default magnitudes and the pseudo-total Kron magnitudes is small there is a
large spread with some galaxies having default magnitudes more than 
0.5~magnitudes fainter than the Kron magnitude. The Kron
magnitudes are systematically fainter than the extrapolated magnitudes by
between approximately 0.1 and 0.3~magnitudes.  This offset is rather larger
than expected: if the Kron radius is computed using a faint isophote to
define the extent of the image from which the intensity weighted radius is
measured, then the Kron magnitudes should be very close to total.  For an
exponential light profile ($\beta=1$), the Kron radius should capture 96\%
of the flux, while for an $r^{1/4}$ law ($\beta=4$), 90\% of the flux
should be enclosed. In other words, the Kron magnitude should differ from
the total magnitude by only 0.044 and 0.11~magnitudes in these two
cases. However, the choice of isophote is a compromise between depth and
statistical robustness. In the case of the \twomass\ second incremental
release, an isophote of 21.7(20.0)~mag arcsec$^{-2}$ in J(\K) was adopted
(Jarrett private communication). These relatively bright isophotes, 
particularly the \K-band  isophote, could
lead to underestimates of the Kron radii and fluxes for 
lower surface brightness objects and plausibly accounts for much of 
the median offset of 0.3~magnitudes seen in Fig.~\ref{fig:mags} between the 
\K-band Kron and extrapolated magnitudes.  This line of reasoning favours 
adopting the extrapolated magnitudes as the best estimate of the total 
magnitudes, but, on the other hand, the extrapolated magnitudes are 
model-dependent and have larger measurement errors.  

To understand better the offset and scatter in the \twomass\
magnitudes we have compared a subset of the \twomass\ data with the
independent K-band photometry of Loveday (2000). The pointed
observations of Loveday have better resolution than the 2MASS images
and good signal-to-noise to a much deeper isophote.
This enables accurate, unbiased Kron magnitudes to be
measured.  Note that the offset between the \twomass\ \K-band and the
standard K-band used by Loveday is expected to be almost completely
negligible (see Carpenter \shortcite{carpenter}).  The left hand
panels of Fig.~\ref{fig:loveday} compare these measurements with the
corresponding \twomass\ Kron and extrapolated magnitudes.  The right
hand panels show \K -band Kron and extrapolated magnitudes computed by
taking the \twomass\ J-band Kron and extrapolated magnitudes and
subtracting the J$-$\K\ colour measured within the default aperture.
These indirect estimates are interesting to consider as they combine
the profile information from the deeper J-band image with the J$-$\K\
colour measured within the largest aperture in which there is good
signal-to-noise.  The straight lines plotted in Fig.~\ref{fig:loveday}
show simple least squares fits and the slope and zero-point offset of
these fits are indicated on each panel along with bootstrap error
estimates. Also shown in the inset panels is the distribution of
residual magnitude differences about each of the fits and a gaussian
fit to this distribution. The rms of these residuals and a bootstrap
error estimate is also given in each panel.

From these comparisons we first see that all the fits have slopes
entirely consistent with unity, but that their zeropoints and scatters
vary. The zero-point offsets, $\CalkDelta$,
between both the \twomass\ Kron magnitude
measurements and those of Loveday confirm that the \twomass\ Kron
magnitudes systematically underestimate the galaxy luminosities. In
the case of the direct \K-band \twomass\ magnitudes the offset is
$\CalkDelta=0.164$~magnitudes.  
In the case of the Kron magnitudes inferred from
the deeper J-band image profiles, the offset is reduced to 
$\CalkDelta=0.061$~magnitudes.
Conversely the \twomass\ extrapolated magnitudes are systematically
brighter than the Loveday Kron magnitudes by 
$-\CaleDelta=0.137$ and~$0.158$~magnitudes, 
where one would expect an offset of only  $\KDelta=0.044$ to $0.11$ 
due to the difference in definition between ideal Kron and true total
magnitudes.  For both estimates of the extrapolated magnitude and for
the directly estimated Kron magnitude the scatter about the
correlation is approximately 0.14~magnitudes and we note a slight
tendency for the scatter to increase at faint magnitudes.  The
magnitude estimate that best correlates with the Loveday measurements
is the Kron magnitude estimated from the \twomass\ J-band Kron
magnitude and the default aperture J$-$\K\ colour. Here the
distribution of residuals has a much reduced scatter of only 
0.1~magnitudes and has very few outliers.

Our conclusions from the comparison of Kron magnitudes is that it is
preferable to adopt the \K-band magnitude inferred from the J-band
Kron or extrapolated magnitude by converting to the \K-band using
default aperture colour, rather than to use the noisier and more biased
direct \K-band estimates.  With this definition, we find that the
\twomass\ Kron magnitudes slightly underestimate the galaxy
luminosities while the extrapolated magnitudes slightly overestimate
the luminosities, particularly at faint fluxes.  We will present
results for both magnitude definitions, but we note that to convert to
total magnitudes we estimate that the \twomass\ Kron magnitudes should
be brightened by $\CalkDelta+\KDelta=0.1$--$0.17$ magnitudes
and the extrapolated
magnitudes dimmed by $-\CaleDelta-\KDelta=0.05$--$0.11$ magnitudes.

\subsection{Completeness of the \twomass\ Catalogue}
\label{sec:compl}

Here we define the magnitude limited samples which we will
analyze in Section~\ref{sec:lf_est} and test them for possible
incompleteness in both magnitude and surface brightness.
For the Kron and extrapolated magnitudes, the \twomass\ catalogue has high
completeness to the nominal limits of J$<$14.7 and \K$<$13.9.  However, to
ensure very high completeness and avoid any bias in our luminosity function
estimates, we made the following more conservative cuts.  For the Kron
magnitudes, we limited our sample to either J$<$14.45 or \K$<$13.2, and for
the extrapolated magnitudes to either J$<$14.15 or \K$<$12.9.  These
choices are motivated by plots such as the top panel of Fig.~\ref{fig:mags}.
Here the isophotal default magnitude limit of
J$<$$14.7$ is responsible for the right hand edge to the distribution of 
data points. One sees that this limit begins to remove objects from the 
distribution of Kron magnitudes for J$\gsim$$14.5$. 
An indication that the survey is complete to our adopted limits is given 
by the number counts shown in Fig~\ref{fig:counts}, which only begin to 
roll over at fainter magnitudes.

\begin{figure}
\centering
\centerline{\epsfxsize=8.5 truecm\epsfbox[0 45 530 755]{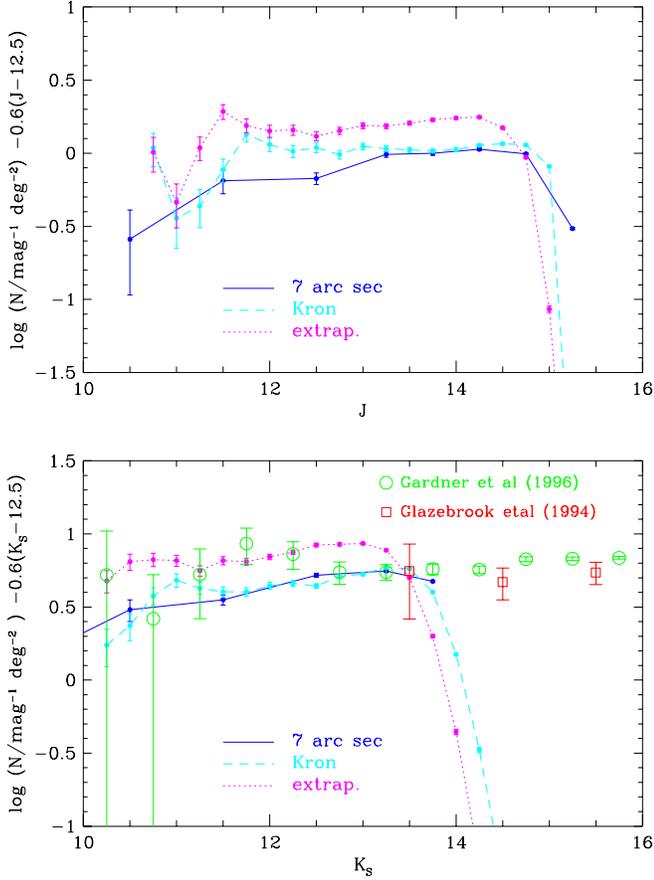}}
\caption{Differential galaxy number counts in the J and \K\ bands,
all with Poisson errorbars and with a Euclidean slope subtracted so
as to expand the scale of the ordinate. The J and \K\ counts linked by the
solid line are the \twomass\ 7~arcsec aperture counts of
Jarrett \etal (in preparation). The counts linked by the
dashed and dotted lines are those of the \twomass--\twodF\  redshift
catalogue for Kron and extrapolated magnitudes respectively.
The \K-band magnitudes are those inferred from the J-band
magnitudes and aperture colours. 
In the \K-band these are compared with the counts of 
Gardner \etal (1996) and Glazebrook \etal (1994) as indicated in the
figure legend.}
\label{fig:counts}
\end{figure}

\begin{figure}
\centering
\centerline{\epsfxsize=8.0 truecm \epsfbox[35 250 550 770]{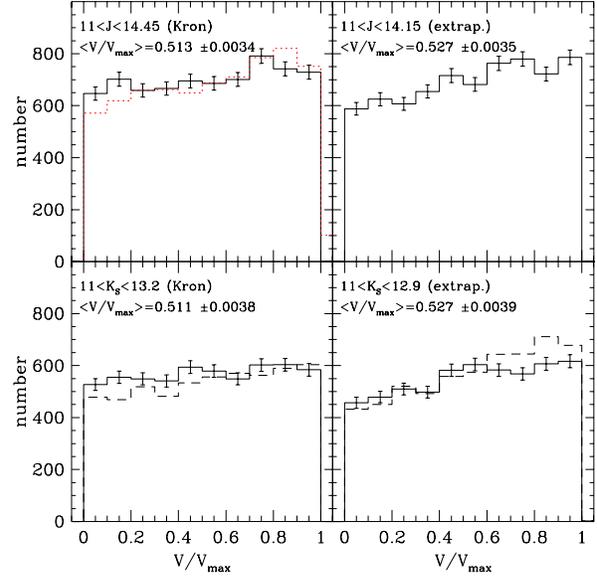}}
\caption{The distributions of $V/V_{\rm max}$ for our magnitude limited
samples. 
The solid histograms in the four panels show the 
$V/V_{\rm max}$ distributions for our J and \K\ Kron and extrapolated
magnitude limited samples. The mean values of
$\langle V/V_{\rm max} \rangle$ are indicated on each panel.
The \K-band magnitudes are those inferred from the J-band
magnitudes and aperture colours. The distributions 
for the directly measured \K-band Kron and extrapolated magnitudes 
are shown by the dashed histograms in the lower panels.
The dotted histogram in the top-left panel shows the 
$V/V_{\rm max}$ distribution we obtain when attempting to
take account of the \twomass\ isophotal diameter and isophotal magnitude 
limits in estimating the $V_{\rm max}$ values.
}
\label{fig:v_vmax}
\end{figure}

More rigorously, we have verified that the
samples are complete to these limits by examining their $V(z_i)/V(z_{{\rm
max},i})$ distributions. Here, $z_i$ is the redshift of a galaxy in the
sample, $z_{{\rm max},i}$ is the maximum redshift at which this galaxy would
satisfy the sample selection criteria, and $V(z)$ is the survey volume that
lies at redshift less than $z$. If the sample is complete and of uniform 
density, $V(z_i)/V(z_{{\rm max},i})$ is uniformly distributed within the 
interval 0 to~1.  To evaluate $z_{\rm max}$  we made use of the default k+e
corrections described in the following section, but the results are
not sensitive to reasonable variations in the assumed corrections or in
the cosmology.
The solid histograms in Fig.~\ref{fig:v_vmax} show these distributions
for each of our four magnitude limited samples. Note that the \K-band 
magnitudes are those inferred from the J-band magnitudes and aperture colours. 
The dashed histograms in the lower panels show the corresponding
distributions for the directly measured \K-band magnitudes. In all these
cases we have computed $V_{\rm max}$ simply from the imposed apparent 
magnitude limits and have ignored any possible dependence of the catalogue 
completeness on surface brightness. 

If the samples were incomplete the symptom one would expect to see
is a deficit in the $V/V_{\rm max}$ distributions at large
$V/V_{\rm max}$ and hence a mean $\langle V/V_{\rm max} \rangle$$<$$0.5$.
There is no evidence for such a deficit in these distributions.
In fact each has a mean $\langle V/V_{\rm max} \rangle$ slightly greater
than $0.5$. The slight gradient in the $V/V_{\rm max}$ distribution
is directly related to the galaxy number counts shown in 
Fig~\ref{fig:counts}, which are slightly steeper than expected
for a homogeneous, non-evolving galaxy distribution. A similar result
has been found in the bright \B -band counts (\cite{apmcounts}). 
The \B-band result has variously been interpreted as evidence for
rapid evolution, systematic errors in the magnitude calibration,
or a local hole or underdensity in the galaxy distribution 
(\cite{apmcounts,metcalfe,shanks}).
Here we note that the gradient in the $V/V_{\rm max}$ distributions
(and also in the galaxy counts) becomes steeper both as one switches
from Kron to the less reliable extrapolated magnitudes and as one switches
from the J-band data to the lower signal-to-noise \K-band data.
This gives strong support to our decision to adopt the \K-band 
magnitudes derived from the J-band Kron and extrapolated magnitudes
and aperture J$-$\K\ colours. It also cautions that the 
mean $\langle V/V_{\rm max} \rangle$$>$$0.5$ cannot necessarily 
be taken as a sign of evolution or a local underdensity, but may instead
be related to the accuracy of the magnitude measurements. The
comparison to the observations of Loveday (\shortcite{love})
shows no evidence for systematic errors in the magnitudes, but does
not constrain the possibility that the distribution of 
magnitude measurement errors may become 
broader or skewed at fainter magnitudes.
Such variations would affect the $V/V_{\rm max}$ distributions and could
produce the observed behaviour.
We conclude by noting that while the shift in the mean
$\langle V/V_{\rm max} \rangle$ is statistically significant, it
is nevertheless quite small for the samples we analyze and
has little effect on the resulting luminosity function
estimates.

\begin{figure}
\centering
\centerline{\epsfxsize=8.0 truecm \epsfbox[30 400 540 750]{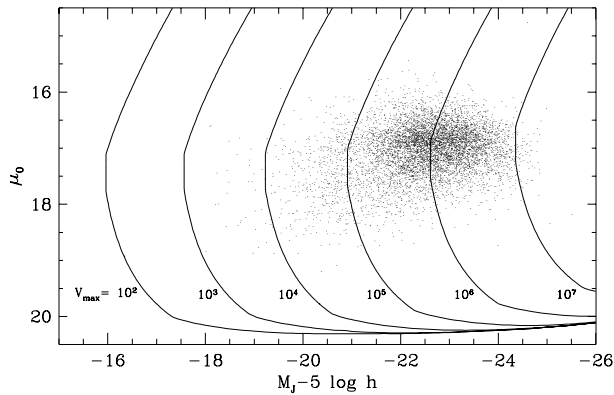}}
\caption{The points show the distribution of estimated central surface
brightness, $\mu_0$, and absolute magnitude, $M_{\rm J}$, for
our  J$<14.45$ (Kron) sample.
The contours show visibility theory estimates of $V_{\rm max}$ as a function of
$\mu_0$ and  $M_{\rm J}$. The contours are labelled
by their $V_{\rm max}$ values in units of (Mpc/h)$^3$.
}
\label{fig:sb}
\end{figure}

We now investigate explicitly the degree to which the
completeness of the \twomass\ catalogue depends on surface brightness
by estimating $V_{\rm max}$ as a function of both absolute magnitude
and surface brightness. This is an important issue: if the catalogue
is missing low-surface brightness galaxies our estimates of the
luminosity function 
will be biased. The approach we have taken follows that developed in 
Cross \etal (\shortcite{cross}) for the \twodF . We estimate an effective 
central surface brightness, $\mu_0^z$, for each observed galaxy assuming 
an exponential light distribution, that the Kron magnitudes are total and 
that the 
Kron radii are exactly five exponential scale-lengths. This is then 
corrected to redshift $z=0$ using
\begin{equation}
	\mu_0=\mu_0^z - 10\log(1+z) - k(z) - e(z)
\end{equation}
to account for redshift dimming and k+e corrections 
(c.f Section~\ref{sec:pop}). Note that in the
\twomass\ catalogue, galaxies with estimated Kron radii less than $7$~arcsec, 
have their Kron radii set to 7 arcsec. 
This will lead us to underestimate the central surface brightnesses of
these galaxies, but this will only affect high surface brightness 
objects and will not affect whether a galaxy can or cannot be seen.
The distribution in the $M_{\rm J}$--$\mu_0$ plane
of our Kron J-band selected sample is shown by the points in 
Fig.~\ref{fig:sb}.

Cross \etal (\shortcite{cross}) use two 
different methods to estimate the value of $V_{\rm max}$  
associated with each position in this plane.
The first method uses the visibility theory of Phillipps, Davies
\& Disney (\shortcite{pbd90}). We model the
selection characteristics of the \twomass\ extended source catalogue
by a set of thresholds. The values appropriate in the J-band are
a minimum isophotal diameter of $8.5$~arcsec at an isophote of 
$20.5$~mag arcsec$^{-2}$, and an isophotal 
magnitude limit of J$<$$14.7$ at an isophote of 
$21.0$~mag arcsec$^{-2}$ (Jarrett et al. 2000). 
In addition, we impose the limits in the Kron magnitude of $11$$<$J$<$$14.45$
that define the sample we analyze. We then calculate for each point
on the $M_{\rm J}$--$\mu_0$ plane the redshift at which a such
a galaxy will drop below one or other of these selection thresholds
and hence compute a value of $V_{\rm max}$. The results of this
procedure are shown by the contours of constant $V_{\rm max}$
plotted in Fig.~\ref{fig:sb}. 
Note that these estimates of $V_{\rm max}$ are only approximate
since we have made the crude assumption that all the galaxies
are circular exponential disks. In addition,
the diameter and isophotal limits
are only approximate and vary with observing conditions.

The second method developed by Cross \etal (\shortcite{cross}) consists
of making an empirical estimate of $V_{\rm max}$ in bins in the
$M_{\rm J}$--$\mu_0$ plane. They look at the distribution 
of observed redshifts in a given bin and adopt the 90$^{\rm th}$ percentile
of this distribution to define $z_{\rm max}$ and hence $V_{\rm max}$.
It is more robust to use the 90$^{\rm th}$ percentile rather than the
100$^{\rm th}$ percentile and the effect of this choice can easily 
be compensated for when estimating the luminosity function
(\cite{cross}). Note that in our application to the \twomass\ data
we do not apply corrections for incompleteness or the effects of 
clustering. The result of this procedure is to confirm that
for the populated bins, the $V_{\rm max}$ values given by the
visibility theory are a good description of the data.

\begin{figure}
\centering
\centerline{\epsfxsize=8.5 truecm \epsfbox[18 390 540 750]{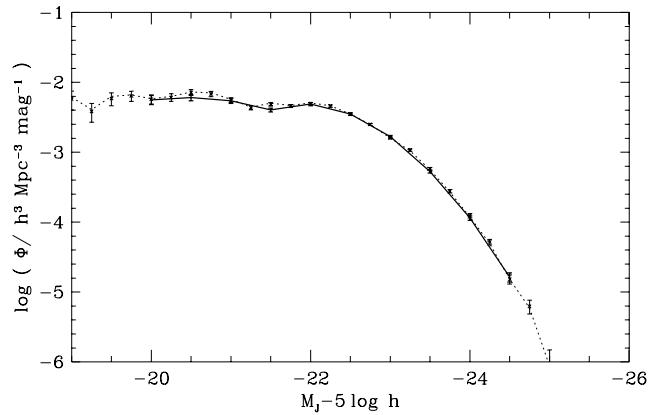}}
\caption{Three $1/V_{\rm max}$ estimates of the Kron J-band 
luminosity function. The data points with errorbars show the estimate 
based on assuming
that $V_{\rm max}$  depends only upon absolute
magnitude and ignoring any possible surface brightness dependence.
The dotted line and heavy solid line show the estimates in which the 
surface brightness
dependence of $V_{\rm max}$ is derived from visibility theory
and from the empirical method of Cross \etal (2000) respectively.
}
\label{fig:lf_sb}
\end{figure}

In Fig.~\ref{fig:sb} we see that
the distribution of galaxies in the $M_{\rm J}$--$\mu_0$ plane
is well separated from the low surface brightness limit
of approximately $20.5$~mag arcsec$^{-2}$ where the 
$V_{\rm max}$ contours indicate that the survey has very little
sensitivity. Thus, there is no evidence that low-surface brightness galaxies
are missing from the \twomass\ catalogue. Furthermore, in the
region occupied by the observed data, the $V_{\rm max}$ contours
are close to vertical indicating that there is little dependence of
$V_{\rm max}$ on surface brightness. The way in which the
$V/V_{\rm max}$ distribution is modified by including this estimate of
the surface brightness dependence is shown by the dotted histogram in 
the top-left panel of Fig.~\ref{fig:v_vmax}. Its effect is to 
increase the mean $V/V_{\rm max}$ slightly, suggesting that this 
estimate perhaps
overcorrects for the effect of surface brightness selection. Even so,
the change in the estimated luminosity function is negligible
as confirmed by the three estimates of the
Kron J-band luminosity function shown in Fig.~\ref{fig:lf_sb}.
These are all simple $1/V_{\rm max}$ estimates, but with 
$V_{\rm max}$ computed either ignoring surface brightness effects or
using one of the two methods described above. These luminosity
functions differ negligibly, indicating that no bias
is introduced by ignoring surface brightness selection effects.

\section{Modelling the Stellar Populations}
\label{sec:pop}

\begin{figure}
\centering
\centerline{\epsfxsize=8.5 truecm \epsfbox[0 420 540 750]{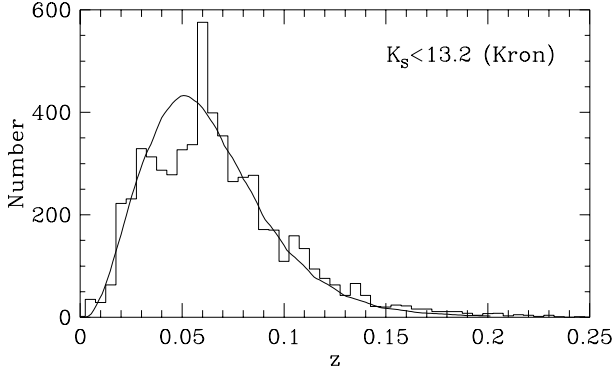}}
\caption{The redshift distribution of the \K$<$$13.2$ (Kron) 
sample selected from the matched \twomass--\twodF\ catalogue. The smooth
curve is the model prediction based on the SWML estimate of the 
\K-band luminosity function (c.f. Section~\ref{sec:lf_est}). 
The model prediction is very insensitive to
the assumed k+e correction and cosmology. 
}
\label{fig:dn_dz}
\end{figure}

\begin{figure}
\centering
\centerline{\epsfxsize=8.5 truecm \epsfbox[90 190 410 706]{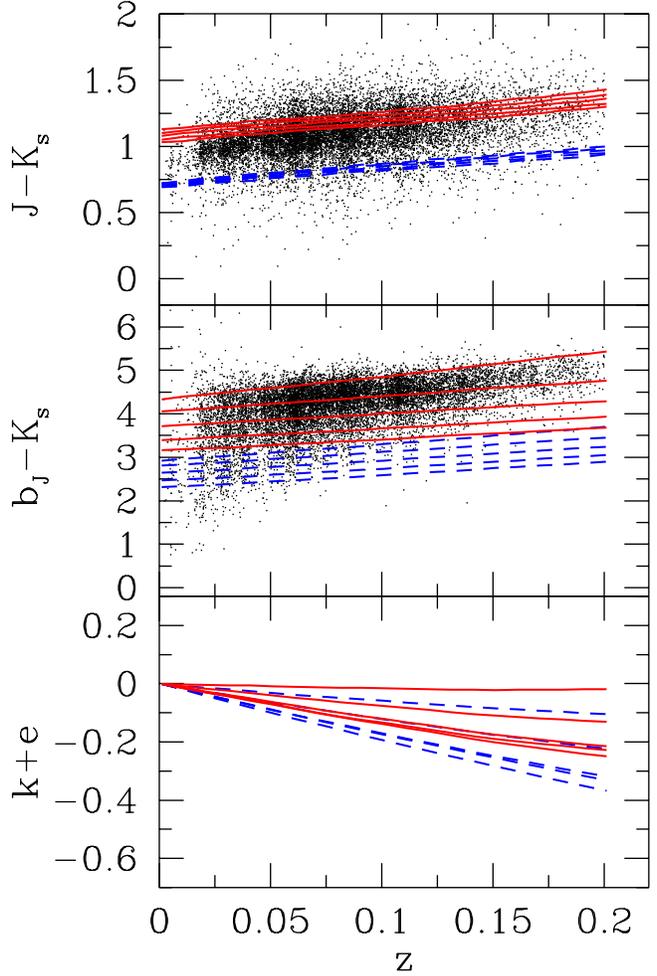}}
\caption{The points in the upper two panels show the observed distributions of
J-\K\ and \B-\K colours as a function of redshift for our matched 
\twomass-\twodF\ catalogue with $z<0.2$ and J$<14.45$ (Kron). 
Overlaid on these points are some examples of model tracks. 
The solid curves are for solar metallicity, $Z=0.02$, and the
dashed curves for $Z=0.004$. 
Within each set, the tracks show different 
choices of the star formation time scale, $\tau$. 
The grid of values we use has $\tau=$1, 3, 5, 10 and 50~Gyr.
Shorter values of $\tau$ lead to older stellar populations and redder 
colours.
The bottom panel shows the k+e corrections in the J-band
for these same sets of tracks.
}
\label{fig:tracks}
\end{figure}

The primary aim of this paper is to determine
the present-day J and \K-band luminosity
functions and also the stellar mass function of galaxies. Since the
\twomass\ survey spans a range of redshift (see Fig.~\ref{fig:dn_dz}), we
must correct for both the redshifting of the filter bandpass 
(k-correction) and for the effects of galaxy evolution (e-correction). 
In practice, the k  and e-corrections at these wavelengths
are both small and uncertainties in them have little
effect on the estimated luminosity functions.  This is because these
infrared bands are not dominated by young stars and also because the
\twomass\ survey does not probe a large range of redshift. We have
chosen to derive individual k and e-corrections for each galaxy using
the stellar population synthesis models of Bruzual \& Charlot
(\shortcite{bc93,bc00}). We have taken this approach not because such
detailed modelling is necessary to derive robust luminosity functions, but
because it enables us to explore the uncertainties in the derived
galaxy stellar mass functions, which are, in fact, dominated by
uncertainties in the properties of the stellar populations.

The latest models of Bruzual \& Charlot (\shortcite{bc00}) provide,
for a variety of different stellar initial mass functions (IMFs), 
the spectral energy distribution (SED), $l_\lambda(t,Z)$, of a single 
population of stars formed at the same time with a single metallicity, 
as a function of both age, $t$, and metallicity, $Z$. We convolve these 
with an assumed star formation history, 
$\psi(t^\prime)$,  to compute the time-evolving SED of the model galaxy,
\begin{equation}
L_\lambda(t) = \int_0^t l_\lambda(t-t^\prime,Z) 
\  \psi(t^\prime) 
\ dt^\prime.
\end{equation}
We take account of the effect of dust extinction on the SEDs using the
Ferrara \etal (\shortcite{fbcg99}) extinction model normalized so that
the V-band central face-on optical depth of the Milk-Way is 10.  This
value corresponds to the mean optical depth of $L_\star$ galaxies in
the model of Cole \etal (\shortcite{cole2000}) which employs the same
model of dust extinction.  We assume a typical inclination angle of 60
degrees which yields a net attenuation factor of 0.53 in the V-band
and 0.78 in the J-band.  By varying the assumed metallicity, $Z$, and
star formation history, we build up a two-dimensional grid of models.
Then, for each of these models, we extract tracks of \B$-$\K\ and
J$-$\K\ colours and stellar mass-to-light ratio as a function of
redshift.

Our standard set of tracks assumes a cosmological model with
$\Omega_0=0.3$, $\Lambda_0=0.7$, Hubble constant $H_0=70$ km
s$^{-1}$ Mpc$^{-1}$, and star formation histories with an
exponential form, $\psi(t) \propto \exp(-[t(z)-t(z_{\rm f})]/\tau)$.
Here, $t(z)$ is the age of the Universe at redshift $z$ and the galaxy
is assumed to start forming stars at $z_{\rm f}=20$ . For
these tracks, we adopt the Kennicutt IMF (\cite{kenn83}) and include
the dust extinction model.  The individual tracks are 
labelled by a metallicity, $Z$,
which varies from $Z=0.0001$ to $Z=0.05$
and a star formation timescale, $\tau$,
which varies from $\tau=1$ Gyr to $\tau=50$ Gyr.
Examples of these tracks are shown in Fig.~\ref{fig:tracks}, along with
the observed redshifts and colours of the \twomass\ galaxies.
We can see that the infrared J$-$\K\ colour depends mainly on metallicity
while the \B$-$\K\ colour depends both on metallicity and star formation
timescale. Thus, the use of both colours allows a unique track to be selected.
Note from the bottom panel that, for all the tracks, the k+e
correction is always small for the range of redshift spanned 
by our data.

We can gauge how robust our results are by varying the assumptions of our
model. In particular, we vary the IMF, the dust extinction and cosmological
models, and include or exclude the evolutionary contribution to the k+e
correction.  Also, we consider power-law star formation histories, $\psi(t)
\propto [t(z)/t(z_{\rm f})]^{-\gamma}$, as an alternative to the
exponential model.  The results are discussed at beginning of
Section~\ref{sec:results}.

The procedure for computing the individual galaxy k+e corrections 
is straightforward. At the measured redshift of a galaxy, 
we find the model whose \B$-$\K\ and J$-$\K\ colours most closely
match that of the observed galaxy. Having selected the model we
then follow it to $z=0$ to predict the galaxy's present-day J and \K-band
luminosities and also its total stellar mass. We also use the
model track to follow its k+e correction to higher redshift in order
to compute $z_{\rm max}$, the maximum redshift at which this galaxy
would have passed the selection criteria  for inclusion into the 
analysis sample.

\section{Luminosity Function Estimation}
\label{sec:lf_est}

We use both the simple $1/V_{\rm max}$ method and
standard maximum likelihood methods to estimate luminosity
functions. We present Schechter function fits computed using the
STY method (Sandage, Tammann \& Yahil \shortcite{sty}) and also
non-parametric estimates using the stepwise maximum likelihood method
(SWML) of Efstathiou, Ellis \& Peterson (\shortcite{eep}). 
Our implementation of each of these methods is described and tested in 
Norberg \etal (\shortcite{norberg00}). The advantage of the
maximum likelihood methods is that they are not affected by 
galaxy clustering (provided that the galaxy luminosity
function is independent of galaxy density). By contrast  the
$1/V_{\rm max}$ method, which makes no assumption about the dependence
of the luminosity function with density, is subject to biases
produced by density fluctuations.

The two maximum likelihood methods determine the shape of the
luminosity function, but not its overall normalization.  We have
chosen to normalize the luminosity functions by matching the galaxy
number counts of Jarrett \etal (in preparation). These were obtained
from a 184~deg$^2$ area selected to have low stellar density and in
which all the galaxy classifications have all been confirmed by
eye. The counts are reproduced in Fig.~\ref{fig:counts}.  By using the
same 7~arcsec aperture magnitudes as Jarrett \etal (in preparation)
and scaling the galaxy counts in our redshift survey, we deduce that
the effective area of our redshift catalogue is
$\areaeff\pm25$~deg$^2$.  Note that normalizing in this way by-passes
the problem of whether or not some fraction of the missed \twomass\
objects are stars.  
Fig.~\ref{fig:counts} also shows the Kron and extrapolated magnitude J
and \K\ counts of the \twomass--\twodF\ redshift survey.  In the lower
panel, these counts are seen to be in agreement with the published
K-band counts of Gardner \etal (\shortcite{gard96}) and Glazebrook
\etal (\shortcite{kgb94}).

We also checked the normalization using the
following independent estimate of the effective solid angle of the
redshift survey.  For galaxies in the \twodF\ parent catalogue
brighter than \B$<B_{\rm limit}$, we computed the fraction that have
both measured redshifts and match a \twomass\ galaxy. For a faint
$B_{\rm limit}$ this fraction is small as the \twodF\ catalogue is
much deeper than the \twomass\ catalogue, but as $B_{\rm limit}$ is
made brighter, the fraction asymptotes to the fraction of the area of
the \twodF\ parent catalogue covered by the joint \twomass--\twodF\
redshift survey. By this method we estimate that the effective area of
our redshift catalogue is $\areaeffa\pm22$~deg$^2$, which is in good
agreement with the estimate from the counts of Jarrett \etal .

It should be noted that for neither of these estimates of the effective
survey area do the quoted uncertainties take account of variations in
the number counts due to large scale structure. To estimate the
expected variation in the galaxy number counts within the combined
\twomass -\twodF\ survey due to large scale structure we constructed an
ensemble of mock catalogues from the $\Lambda$CDM Hubble volume
simulation of the VIRGO consortium (Evrard \shortcite{evrard98};
Evrard \etal in preparation;
http://www.physics.lsa.umich.edu/hubble-volume). Mock \twodF\
catalogues constructed from the VIRGO Hubble Volume simulations 
(Baugh \etal in preparation) can be found at 
http://star-www.dur.ac.uk/$\tilde{\hphantom{n}}$cole/mocks/hubble.html .  
We simply took these catalogues and sampled them to the depth of
\twomass\ over a solid angle of $\areaeff$~deg$^2$. To this magnitude
limit we found an rms variation in the number of galaxies of 15\%. We
took this to be a realistic estimate of the uncertainty in the
\twomass\ number counts and propagated this error through when
computing the error on the normalization of the luminosity function.

\section{Results}
\label{sec:results}

\subsection{Luminosity Functions}

\begin{figure*}
\centering
\centerline{\epsfbox[0 520 540 750]{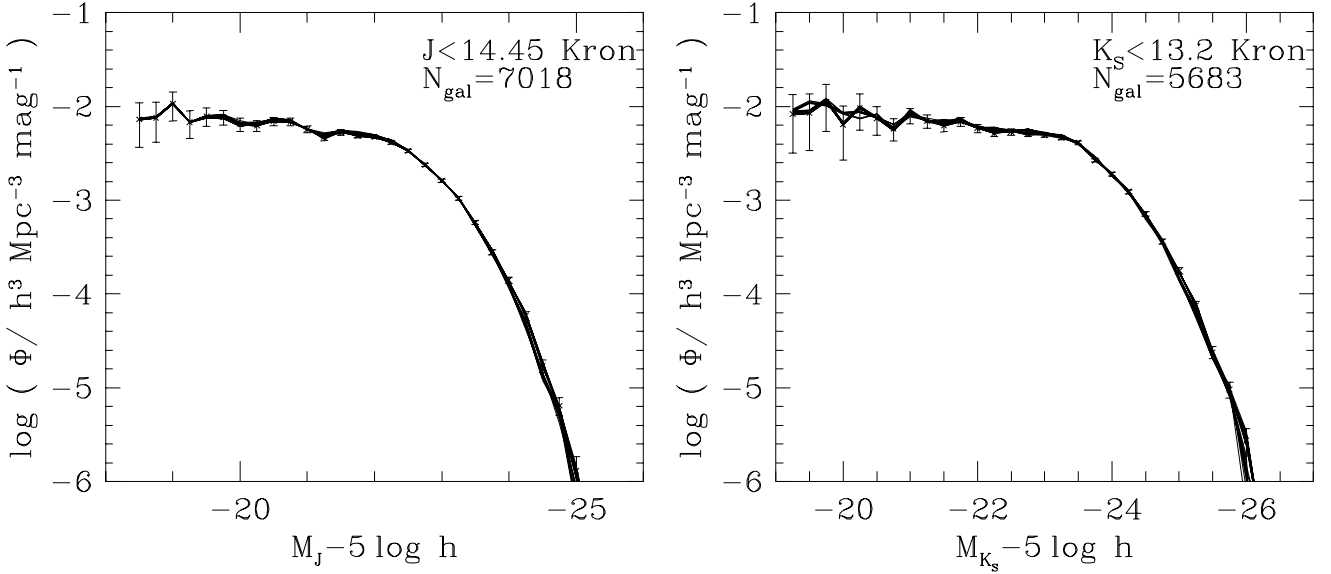}}
\caption{SWML estimates of the 
Kron magnitude J (left) and \K-band (right) luminosity functions
(points with error bars). Our default
model of  k+e corrections (Kennicutt IMF and standard dust extinction)
is adopted. The set of curves on each plot shows the effects of neglecting
dust extinction and/or switching to a Salpeter IMF and/or changing the Hubble 
constant to $H_0=50$ km s$^{-1}$Mpc$^{-1}$ and/or adopting power-law star formation 
histories and/or making a k-correction but no evolution correction.
}
\label{fig:lf_jk1}
\end{figure*}

\begin{table*}
\centering
\caption{The dependence of the J and \K-band
Schechter function parameters on cosmological parameters and evolutionary
corrections.  The parameters refer to STY estimates of 
the luminosity function for the
\twomass\ Kron magnitudes and are derived using k or k+e
corrections based on model tracks that include dust extinction and assume
the Kennicutt IMF. To convert to total magnitudes we estimate that the
$\Mstar$ values should be brightened by between 
$\KDelta+\CalkDelta-\ConvDelta=0.08$ and~$0.15$ magnitudes.
Note that the statistical errors we quote for the $\Phi_\star$ values
include the significant contribution that we estimate is induced by
large scale structure.
}
\begin{center}
\begin{tabular}{lrrrrrrrrrrrrrrrrrr} 
\multicolumn{1}{l} {$\Omega_0$} &
\multicolumn{1}{l} {$\Lambda_0$} &
\multicolumn{1}{l} {Model Tracks} &
\multicolumn{1}{l} {$\Mstarj - 5 \log h$} &
\multicolumn{1}{l} {$\alphaj$} &
\multicolumn{1}{l} {$\phij$/h$^3$Mpc$^{-3}$} &
\multicolumn{1}{l} {$\Mstark - 5 \log h$} &
\multicolumn{1}{l} {$\alphak$} &
\multicolumn{1}{l} {$\phik$/h$^3$Mpc$^{-3}$} & \\
\hline 
0.3 & 0.7 & k$+$e    & -22.36$\pm$0.02 &-0.93$\pm$0.04 &$1.04\pm0.16 \times 10^{-2}$ 
                   & -23.44$\pm$0.03 &-0.96$\pm$0.05 &$1.08\pm0.16 \times 10^{-2}$ \\
0.3 & 0.7 & k only & -22.47$\pm$0.02 &-0.99$\pm$0.04 &$0.90\pm0.14 \times 10^{-2}$ 
                   & -23.51$\pm$0.03 &-1.00$\pm$0.04 &$0.98\pm0.15 \times 10^{-2}$ \\
0.3 & 0.0 & k$+$e    & -22.29$\pm$0.03 &-0.89$\pm$0.04 &$1.16\pm0.18 \times 10^{-2}$ 
                   & -23.36$\pm$0.03 &-0.93$\pm$0.05 &$1.21\pm0.18 \times 10^{-2}$ \\
0.3 & 0.0 & k only & -22.38$\pm$0.03 &-0.95$\pm$0.04 &$1.02\pm0.15 \times 10^{-2}$ 
                   & -23.43$\pm$0.03 &-0.96$\pm$0.05 &$1.10\pm0.16 \times 10^{-2}$ \\
1.0 & 0.0 & k$+$e    & -22.22$\pm$0.02 &-0.87$\pm$0.03 &$1.26\pm0.19 \times 10^{-2}$ 
                   & -23.28$\pm$0.03 &-0.89$\pm$0.05 &$1.34\pm0.20 \times 10^{-2}$ \\
1.0 & 0.0 & k only & -22.34$\pm$0.02 &-0.93$\pm$0.04 &$1.08\pm0.16 \times 10^{-2}$ 
                   & -23.38$\pm$0.03 &-0.93$\pm$0.05 &$1.18\pm0.17\times 10^{-2}$ \\
\hline			        
\end{tabular} 			 
\end{center}
\label{tab:lfpar}
\end{table*}

\begin{table}
\centering
\caption{The SWML J and \K-band luminosity functions 
for Kron magnitudes as plotted in Fig.~\ref{fig:lf_jk2}. 
The units of both $\phi$ and its uncertainty $\Delta \phi$  are number
per $h^{-3}$  Mpc$^3$ per magnitude.}
\begin{center}
\begin{tabular}{llllll} 
\multicolumn{1}{l} {$M - 5 \log h$} &
\multicolumn{1}{l} {$\phi_J \pm \Delta \phi_J$} &
\multicolumn{1}{l} {$\phi_{K_{\rm S}} \pm \Delta \phi_{K_{\rm S}}$}\\ 
\hline 
-18.00 & (5.73$\pm$3.58)$\times 10^{-3}$& (3.13$\pm$3.64)$\times 10^{-3}$\\  
-18.25 & (5.38$\pm$3.34)$\times 10^{-3}$& (8.26$\pm$6.68)$\times 10^{-3}$\\
-18.50 & (7.60$\pm$3.75)$\times 10^{-3}$&                              \\
-18.75 & (7.94$\pm$3.59)$\times 10^{-3}$& (4.65$\pm$4.10)$\times 10^{-3}$\\
-19.00 & (1.11$\pm$3.82)$\times 10^{-2}$& (5.76$\pm$4.32)$\times 10^{-3}$\\
-19.25 & (6.98$\pm$2.26)$\times 10^{-3}$& (9.16$\pm$5.67)$\times 10^{-3}$\\
-19.50 & (8.14$\pm$1.80)$\times 10^{-3}$& (1.12$\pm$0.64)$\times 10^{-2}$\\
-19.75 & (8.17$\pm$1.45)$\times 10^{-3}$& (1.05$\pm$0.57)$\times 10^{-2}$\\
-20.00 & (7.16$\pm$1.12)$\times 10^{-3}$& (8.58$\pm$4.63)$\times 10^{-3}$\\
-20.25 & (6.62$\pm$0.88)$\times 10^{-3}$& (8.82$\pm$3.86)$\times 10^{-3}$\\
-20.50 & (7.30$\pm$0.76)$\times 10^{-3}$& (6.94$\pm$2.44)$\times 10^{-3}$\\
-20.75 & (7.07$\pm$0.64)$\times 10^{-3}$& (6.09$\pm$1.63)$\times 10^{-3}$\\
-21.00 & (5.84$\pm$0.48)$\times 10^{-3}$& (9.26$\pm$1.69)$\times 10^{-3}$\\
-21.25 & (4.97$\pm$0.39)$\times 10^{-3}$& (6.96$\pm$1.18)$\times 10^{-3}$\\
-21.50 & (5.69$\pm$0.35)$\times 10^{-3}$& (7.29$\pm$0.98)$\times 10^{-3}$\\
-21.75 & (5.15$\pm$0.28)$\times 10^{-3}$& (6.99$\pm$0.79)$\times 10^{-3}$\\
-22.00 & (4.89$\pm$0.21)$\times 10^{-3}$& (5.98$\pm$0.61)$\times 10^{-3}$\\
-22.25 & (4.49$\pm$0.17)$\times 10^{-3}$& (5.93$\pm$0.52)$\times 10^{-3}$\\
-22.50 & (3.41$\pm$0.12)$\times 10^{-3}$& (5.39$\pm$0.42)$\times 10^{-3}$\\
-22.75 & (2.37$\pm$0.09)$\times 10^{-3}$& (5.85$\pm$0.37)$\times 10^{-3}$\\
-23.00 & (1.59$\pm$0.06)$\times 10^{-3}$& (5.24$\pm$0.28)$\times 10^{-3}$\\
-23.25 & (1.06$\pm$0.04)$\times 10^{-3}$& (4.96$\pm$0.22)$\times 10^{-3}$\\
-23.50 & (5.41$\pm$0.27)$\times 10^{-4}$& (4.18$\pm$0.17)$\times 10^{-3}$\\
-23.75 & (2.66$\pm$0.17)$\times 10^{-4}$& (2.72$\pm$0.11)$\times 10^{-3}$\\
-24.00 & (1.19$\pm$0.10)$\times 10^{-4}$& (1.88$\pm$0.08)$\times 10^{-3}$\\
-24.25 & (4.69$\pm$0.54)$\times 10^{-5}$& (1.21$\pm$0.06)$\times 10^{-3}$\\
-24.50 & (1.20$\pm$0.22)$\times 10^{-5}$& (6.54$\pm$0.37)$\times 10^{-4}$\\
-24.75 & (5.40$\pm$1.34)$\times 10^{-6}$& (3.46$\pm$0.23)$\times 10^{-4}$\\
-25.00 & (5.42$\pm$3.88)$\times 10^{-7}$& (1.48$\pm$0.13)$\times 10^{-4}$\\
-25.25 &                              &	  (5.55$\pm$0.65)$\times 10^{-5}$\\
-25.50 &                              &	  (2.13$\pm$0.33)$\times 10^{-5}$\\
-25.75 & 	                      &   (9.42$\pm$1.96)$\times 10^{-6}$\\
-26.00 & 	                      &   (1.09$\pm$0.56)$\times 10^{-6}$\\
\hline			        
\end{tabular} 			 
\end{center}
\label{tab:lfswml}
\end{table}

Fig.~\ref{fig:lf_jk1} shows SWML estimates of the Kron  J and
\K\ luminosity functions.  The points with errorbars show 
results for our default choice of k+e corrections, namely those obtained for an
$\Omega_0=0.3$, $\Lambda_0=0.7$, $H_0=70$ km s$^{-1}$
Mpc$^{-1}$ cosmology with a Kennicutt IMF and including dust
extinction. The figure also illustrates that the luminosity functions
are very robust to varying this set of assumptions. The various curves
in each plot are estimates made neglecting dust extinction and/or 
switching to a Salpeter IMF and/or changing the Hubble constant to
$H_0=50$ km s$^{-1}$Mpc$^{-1}$ and/or adopting
power-law star formation histories and/or making a k-correction 
but no evolution correction. The systematic shifts caused by varying
these assumptions are all comparable with 
or smaller than the statistical errors.
The biggest shift results from applying or neglecting
the evolutionary correction. In terms of the characteristic luminosity
in the STY Schechter function fit, the estimates which include
evolutionary corrections are $0.05$ to $0.1$ magnitudes fainter 
than those that only include k-corrections (see Table~\ref{tab:lfpar}).

In Fig.~\ref{fig:lf_jk2} we compare $1/V_{\rm max}$ and SWML Kron
luminosity function estimates (for our default choice of k+e
corrections) with STY Schechter function estimates.  In general, the
luminosity functions are well fit by Schechter functions, but there is
marginal evidence for an excess of very luminous galaxies over that
expected from the fitted Schechter functions. We tabulate the SWML
estimates in Table~\ref{tab:lfswml}. Integrating over
the luminosity function gives luminosity densities in the J and
\K-bands of $\rho_J=(2.75\pm 0.41) \times 10^8 h \Lsun \Mpc^{-3}$ and
$\rho_{K_S}=(5.74\pm 0.86) \times 10^8 h \Lsun \Mpc^{-3}$ respectively,
where we have adopted $M^{\odot}_J=3.73$ and $M^\odot_{K_S}=3.39$
(Allen \shortcite{allen}; Johnson  \shortcite{johnson}).
In this analysis, we have not taken account of the systematic and
random measurement errors in the galaxy magnitudes. In the case of the
STY estimate, the random measurement errors can be accounted for by
fitting a Schechter function which has been convolved with the
distribution of magnitude errors.  However, for the Kron magnitudes,
the rms measurement error is only 0.1~magnitudes, as indicated by the
comparison in the top right hand panel of Fig~\ref{fig:loveday}, and
such a convolution has only a small effect on the resulting Schechter
function parameters.  We find that the only parameter that is affected
is ${M^\star}$ which becomes fainter by just $\ConvDelta=0.02$
magnitudes.  The comparison to the Loveday (\shortcite{love}) data
also indicates a systematic error in the \twomass\ Kron magnitudes of
$\CalkDelta=0.061\pm0.031$. Combining these two systematic errors
results in a net brightening of ${M^\star}$ by
$\CalkDelta-\ConvDelta=0.041\pm0.031$ magnitudes.  As this net
systematic error is both small and uncertain we have chosen not to
apply a correction to our quoted Kron magnitude luminosity function
parameters.  We recall also that to convert from Kron to total magnitudes
requires brightening ${M^\star}$ by between $\KDelta=0.044$ and $0.11$ 
depending on whether the luminosity profile of a typical galaxy is fit 
well by an exponential or $r^{1/4}$-law.

\begin{figure*}
\centering
\centerline{\epsfbox[0 520 540 750]{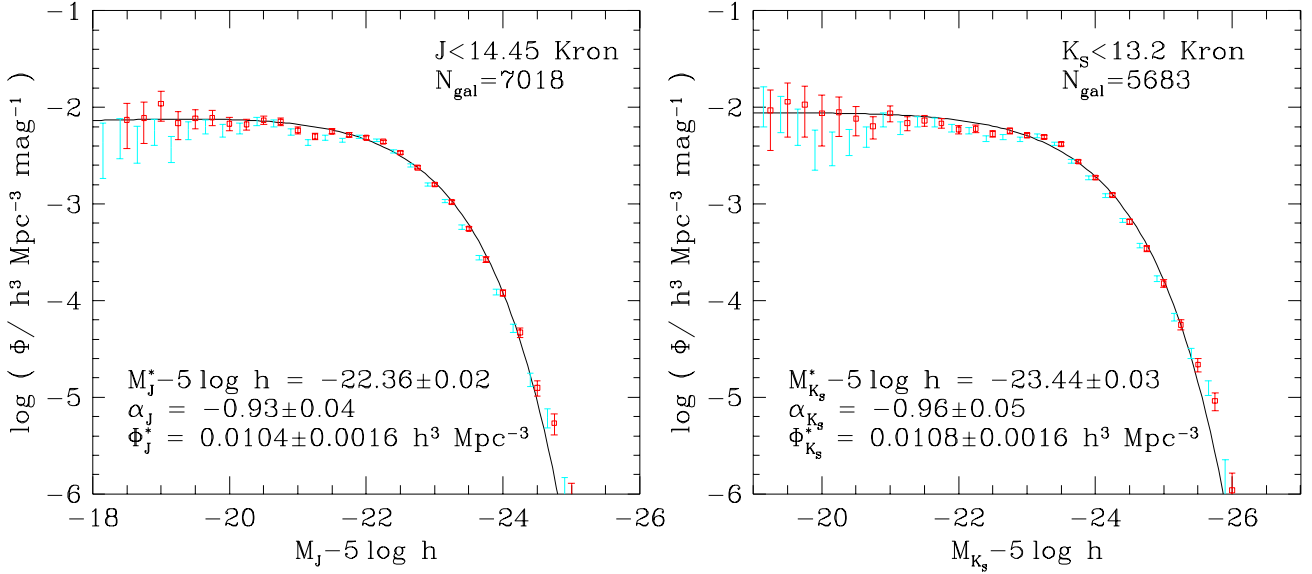}}
\caption{SWML estimates of the Kron
magnitude J (left) and \K-band (right) luminosity functions 
(data points with errorbars)
and STY Schechter function estimates (lines). 
The parameter values and
error estimates of the Schechter functions are given in the legends. 
The errorbars without data points show  $1/V_{\rm max}$ estimates
of the luminosity functions. 
For clarity these have been displaced to the left by 0.1~magnitudes.
}
\label{fig:lf_jk2}
\end{figure*}

\begin{figure*}
\centering
\centerline{\epsfbox[0 520 540 750]{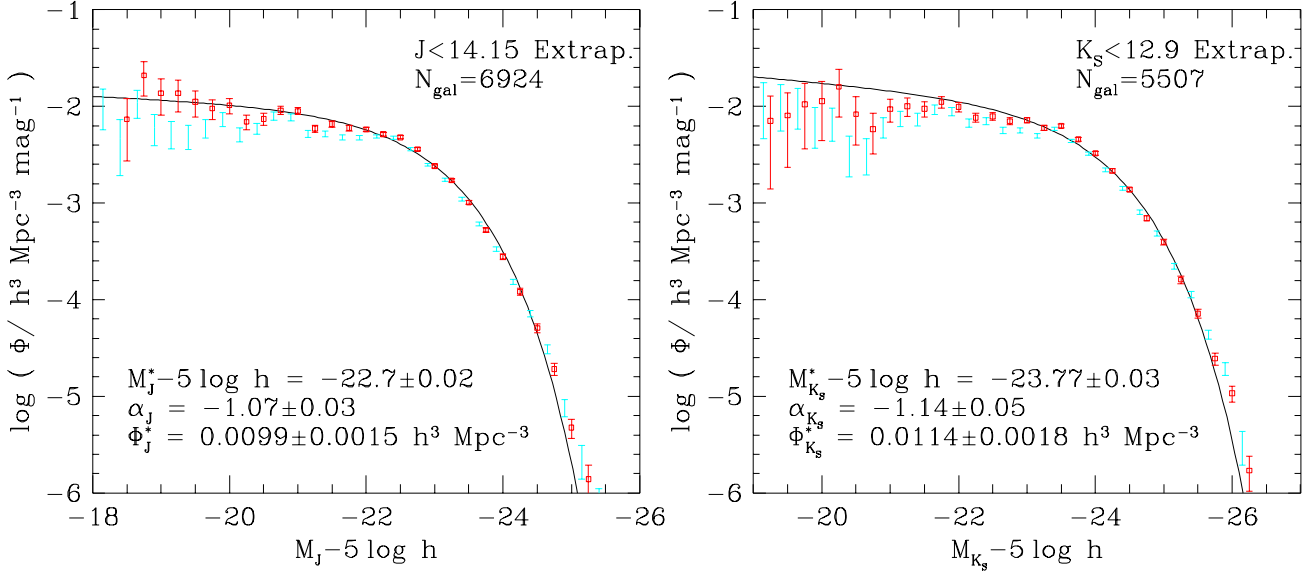}}
\caption{SWML estimates of the 
extrapolated magnitude J (left) and \K-band (right) luminosity functions 
(data points with errorbars)
and STY Schechter function estimates (lines). The parameter values and
error estimates of the Schechter functions are given
in the legends. 
The errorbars without data points show  $1/V_{\rm max}$ estimates
of the luminosity functions. 
For clarity these have been displaced to the left by 0.1~magnitudes.
}
\label{fig:lf_jk3}
\end{figure*}

Fig.~\ref{fig:lf_jk3} shows the SWML and STY luminosity function
estimates for samples defined by the \twomass\ extrapolated, rather
than Kron, magnitudes.  With this definition of magnitude, the
luminosity functions differ significantly from those estimated using
Kron magnitudes.  In particular, the characteristic luminosities are
$0.34$ and $0.28$~magnitudes brighter in J and \K\ respectively.  Most
of this difference is directly related to the systematic offset in the
J-band Kron and extrapolated magnitudes, which can be seen in either
the middle panel of Fig.~\ref{fig:mags} or the right hand panels of
Fig.~\ref{fig:loveday} to be approximately $0.23$~magnitudes.  Note
that even in the \K-band, it is this J-band offset that is relevant as
the \K-band magnitudes we use are derived from the J-band values using
the measured aperture colours. We have argued in
Section~\ref{sec:mags} that this offset is caused by the J-band Kron
\twomass\ magnitudes being fainter than true total magnitudes by
between $\CalkDelta+\KDelta=0.1$ and~$0.17$ and the extrapolated
magnitudes being systematically too bright by
$-\CaleDelta-\KDelta=0.05$ to~$0.11$ magnitudes.  Subtracting this
$0.23$~magnitude offset results in Kron and extrapolated luminosity
functions that differ in $M^\star$ by only 0.11 magnitudes. In the
\K-band, the faint end slope of the best-fit Schechter function is
significantly steeper in the extrapolated magnitude case, but note
that this function is not a good description of the faint end of the
luminosity function since the SWML and $1/V_{\rm max}$ estimates lie
systematically below it. The Schechter function fit is constrained
mainly around $M_\star$ and in this case the $\chi^2$ value indicates
it is not a good fit overall.

The residual differences between the Kron and extrapolated magnitude
luminosity functions arise from the scatter in the relation between
extrapolated and Kron magnitudes.  If this scatter is dominated by
measurement error, then these differences represent small biases,
which are largest for the less robust, extrapolated
magnitudes. However, it is possible that the scatter is due to genuine
variations in galaxy morphology and light profiles.  To assess which
of these alternatives is correct requires independent deep photometry
of a sample of \twomass\ galaxies to quantify the accuracy of the
extrapolated magnitudes. However, we note that the $V/V_{\rm max}$
distributions for the extrapolated magnitudes shown in
Fig.~\ref{fig:v_vmax} have mean $\langle V/V_{\rm max}\rangle$ values
significantly greater than $0.5$, which is probably an indication that
the extrapolated magnitudes are not robust. Thus, overall we favour
adopting Kron magnitudes, noting the small offset of
$\CalkDelta+\KDelta-\ConvDelta=0.08$ to~$0.15$ required to convert to
total magnitudes and correct for the convolving effect of measurement
errors.

The parameters of the STY Schechter function fits shown in
Fig.~\ref{fig:lf_jk2} are listed in the first row of
Table~\ref{tab:lfpar}. The subsequent rows illustrate how the best-fit
parameters change when the cosmological model is varied and the
evolutionary correction is included or excluded.  The $M^\star$ values
are approximately $0.14$ magnitudes fainter for the $\Omega_0=1$ case
than for our standard $\Omega_0=0.3$, $\Lambda_0=0.7$ cosmology.  This
shift is largely due to the difference in distance
moduli between the two cosmologies at the median redshift of the
survey. This, and the difference in the volume-redshift relation, 
cause $\phi_\star$ to change in order to preserve the same galaxy
number counts.

\begin{figure*}
\centering
\centerline{\epsfxsize=15.0 truecm \epsfbox[24 50 535 750]{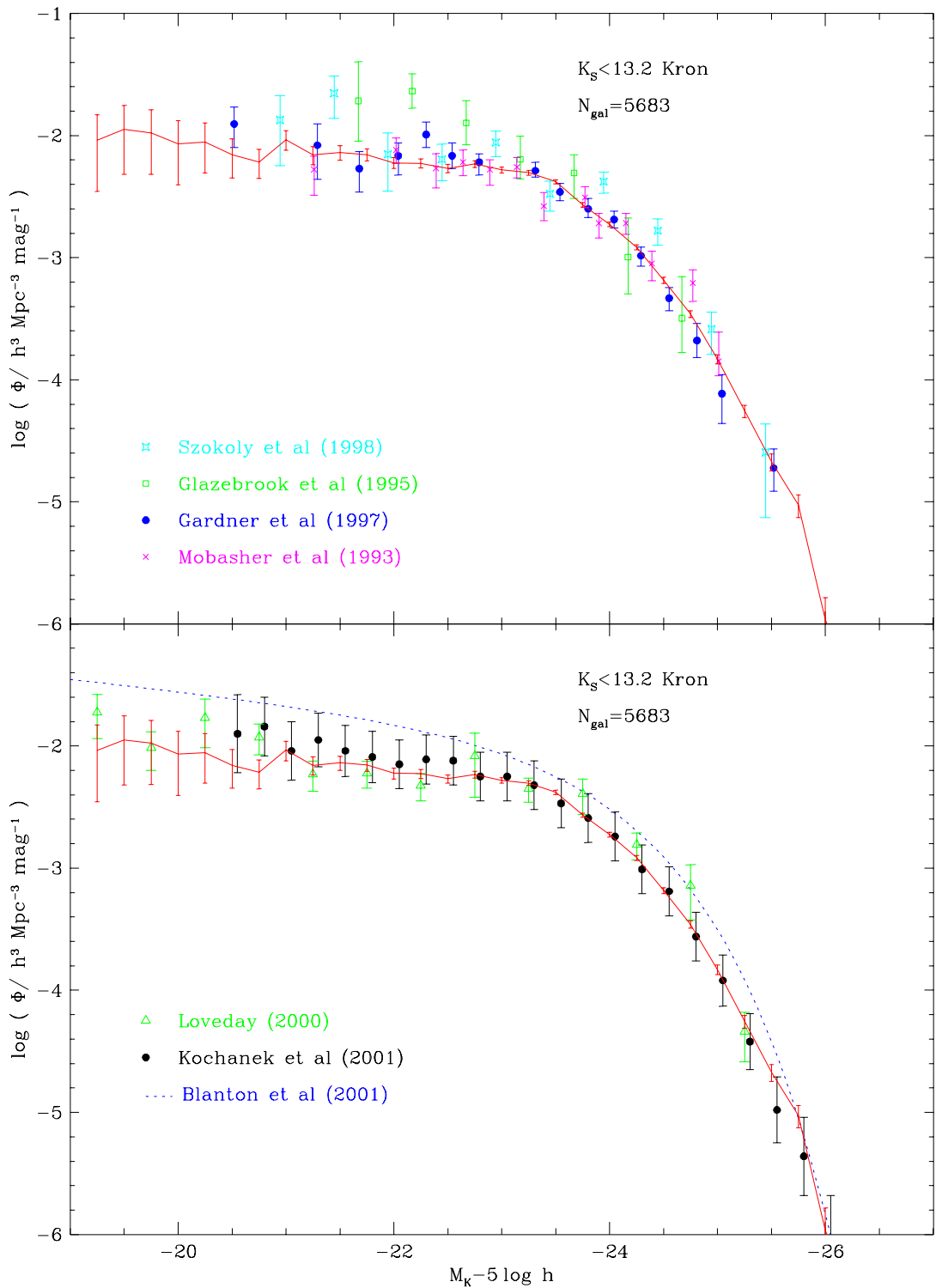}}
\caption{Comparison of various estimates of the K-band luminosity function.
In both panels the solid line shows our SWML estimate of the \K-band 
luminosity function for Kron magnitudes.
The symbols and errorbars in the top panel show the estimates of 
Mobasher \etal (1993), Glazebrook \etal (1995),  Gardner \etal (1997) and
Szokoly \etal (1998) as indicated in the legend. We have shifted the
data of Glazebrook \etal (1995) brightward by 0.3~magnitudes and the
data of Mobasher \etal (1993) faintward by 0.22~magnitudes as
advocated by Glazebrook \etal (1995) to make aperture corrections
and consistent k-corrections respectively.
In the lower panel the symbols and errorbars show the recent estimates of
Loveday (2000) and Kochanek \etal (2000). The estimate of Kochanek \etal
has been shifted brightward by 0.05~magnitudes to account for the difference
between isophotal and Kron magnitudes.
The dotted line shows
a Schechter function estimate of the K-band luminosity function
inferred  from SDSS z$^*$-band luminosity function of Blanton \etal (2000) 
(see text for details). 
}
\label{fig:lf_k4}
\end{figure*}

\begin{table*}
\centering
\caption{Schechter function fits to K-band luminosity functions.
Where necessary, the values quoted have been converted from
the cosmological model assumed in the original work to the
$\Omega_0=0.3$, $\Lambda_0=0.7$ assumed here. In addition, 
we have shifted the $M^\star_{\rm K}$  of Kochanek \etal (2000) 
brightward by 0.05~magnitudes corresponding to the mean difference
between \twomass\ Kron and isophotal magnitudes in the 
Kochanek \etal sample. We have also shifted $M^\star_{\rm K}$ 
of Glazebrook \etal (1995) brightward by 0.3~magnitudes and that
of Mobasher \etal (1993) faintward by 0.22~magnitudes as
advocated by Glazebrook \etal (1995) to make aperture corrections
and consistent k-corrections respectively. 
}
\begin{center}
\begin{tabular}{lrrrrrrrrrrrrrrrrrr} 
\multicolumn{1}{l} {Sample}\hfill &
\multicolumn{1}{l} {$M^\star_{\rm K}$} &
\multicolumn{1}{l} {$\alpha_{\rm K}$} &
\multicolumn{1}{l} {$\Phi_{\rm K}$/h$^3$Mpc$^{-3}$} & \\ 
\hline 
Mobasher et al. 1993  & $-23.37\pm0.30$  & $1.0\hphantom{0}\pm 0.3\hphantom{0}$   & $1.12\pm0.16\times 10^{-2}$ \\
Glazebrook et al. 1995 & $-23.14\pm0.23$ & $1.04\pm 0.3\hphantom{0}$  & $2.22\pm0.53\times 10^{-2}$ \\
Gardner et al. 1997 & $-23.30\pm0.17$    & $1.0\hphantom{0}\pm 0.24$  & $1.44\pm0.20\times 10^{-2}$ \\
Szokoly et al. 1998 & $-23.80 \pm 0.30$   & $1.3\hphantom{0}\pm 0.2\hphantom{0}$   & $0.86\pm0.29\times 10^{-2}$ \\
Loveday et al. 2000 & $-23.58\pm 0.42$   & $1.16\pm 0.19$ & $1.20\pm0.08 \times 10^{-2}$ \\
Kochanek et al. 2000 & $-23.43\pm 0.05$   & $1.09\pm 0.06$ & $1.16\pm0.1\hphantom{0} \times 10^{-2}$ \\

This paper           & $-23.44 \pm 0.03$  & $0.96\pm 0.05$ & $1.08\pm 0.16\times 10^{-2}$ \\
\hline			        
\end{tabular} 			 
\end{center}
\label{tab:lfpar_old}
\end{table*}

\begin{figure*}
\centering
\centerline{ \epsfxsize = 9 truecm
\epsfbox[65 180 440 690]{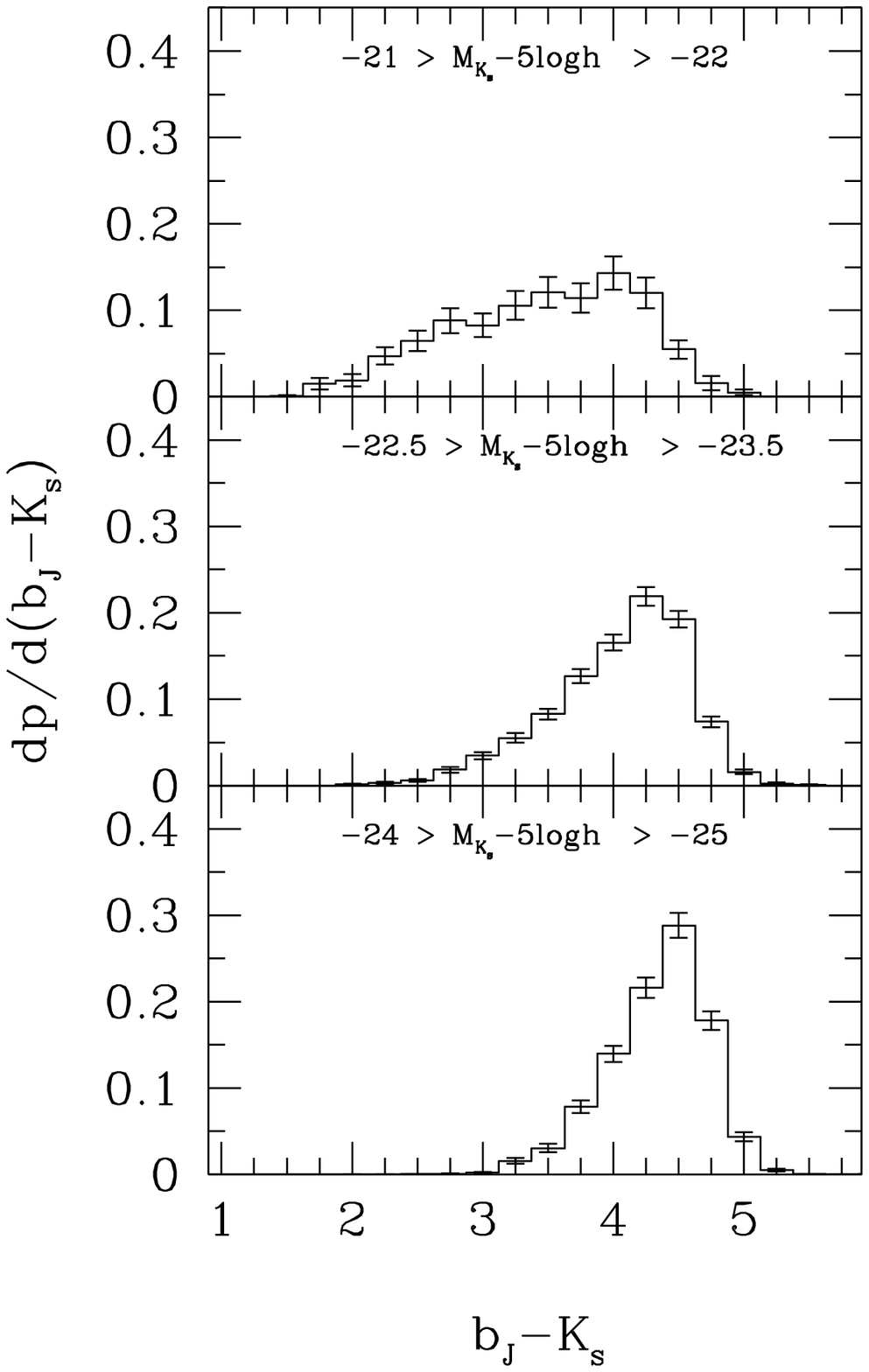}
\epsfxsize = 9 truecm
\epsfbox[65 180 440 690]{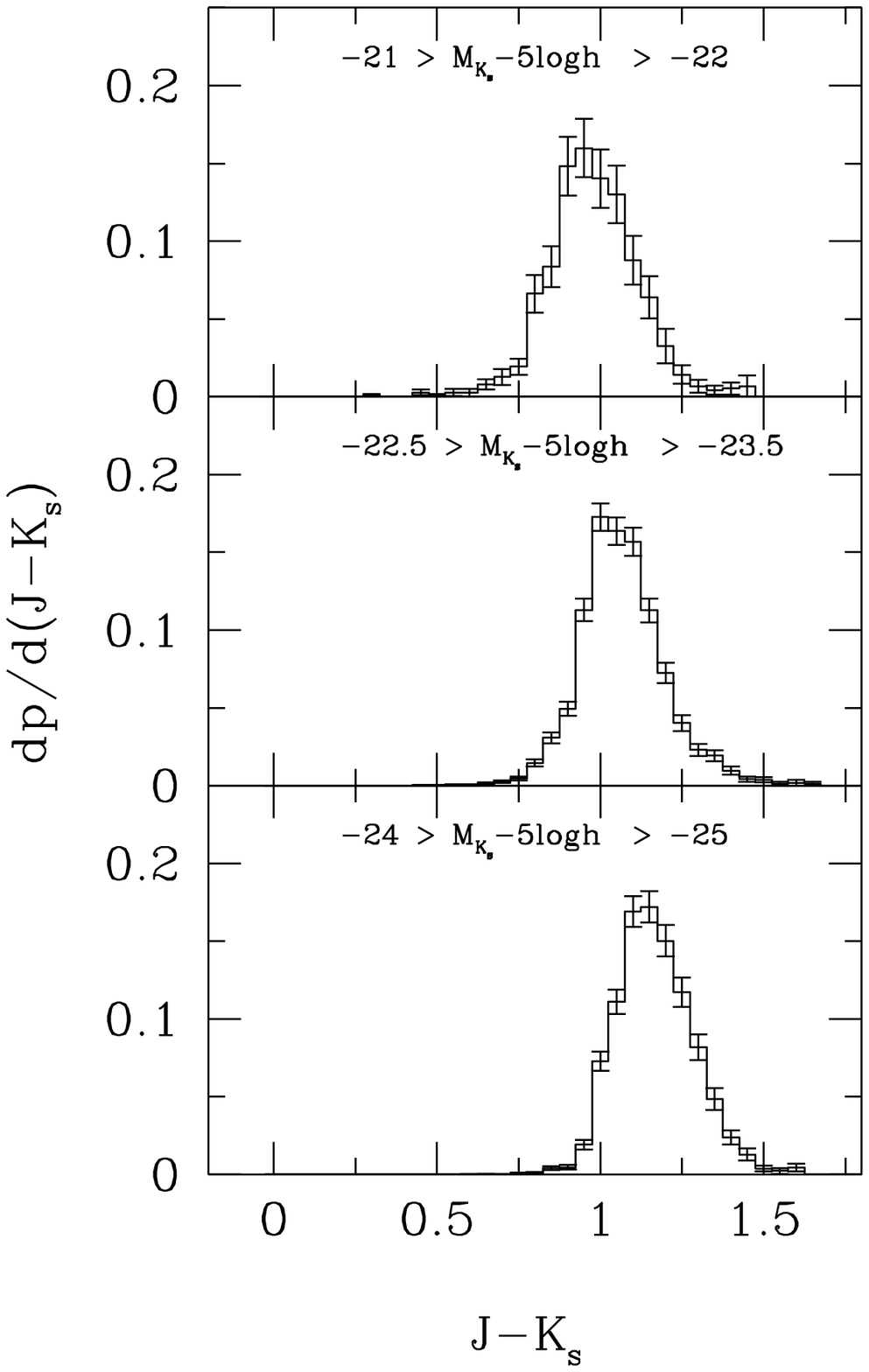}}
\caption{The distribution of rest frame \B$-$\K\ (left) 
and J$-$\K\ (right) colours in three bins of \K\ absolute magnitude,
computed using our default set of k+e corrections.
}
\label{fig:col}
\end{figure*}

Fig.~\ref{fig:lf_k4} compares our estimates of the \K-band luminosity
function for our standard $\Omega_0=0.3$, $\Lambda_0=0.7$ cosmology, with the
estimates of Mobasher \etal (1993), Glazebrook \etal (1995), Gardner \etal
(1997), Szokoly \etal (1998), Loveday (2000) and Kochanek \etal
(\shortcite{kochanek2001}).  In general, these authors
assumed different cosmological models when analysing their data. We have
therefore modified the estimates from each survey.  First, we apply a shift 
in magnitude reflecting the difference in distance moduli, at the median 
redshift, between the assumed cosmological model and  our
standard $\Omega_0=0.3$, $\Lambda_0=0.7$ model. We then apply a 
shift in number density
so as to keep fixed the surface density of galaxies per square degree
at the survey magnitude limit. In the case of the 
Kochanek \etal (\shortcite{kochanek2001}) luminosity function we have
shifted the data points brightwards by 0.05~magnitudes to account
for the mean difference between the \twomass\ isophotal magnitudes used by 
Kochanek \etal (\shortcite{kochanek2001}) and the Kron magnitudes we 
have adopted.
Schechter function parameters scaled and adjusted in this
manner are given for each survey in  Table~\ref{tab:lfpar_old}.
Note that due to the correlations between the Schechter function
parameters it is better to judge the agreement between the different 
estimates by reference to Fig.~\ref{fig:lf_k4} rather
than by the parameter values in Table~\ref{tab:lfpar_old}.
Our new estimate of the \K-band luminosity function is in excellent
agreement with the independent estimates and has the smallest
statistical errors at all magnitudes brighter than $M_{\rm {K_S}} - 5 \log
h = -22$.  For very faint magnitudes, from $-20$ to $-16$, the sparsely
sampled survey of Loveday (2000) has smaller statistical errors.  Note
that many previous analyses of the K-band luminosity function ignored
the contribution of large scale structure to the error in $\Phi_\star$,
and so the errors in Table~\ref{tab:lfpar_old}
are likely to be underestimated.

Also shown on the lower panel of Fig.~\ref{fig:lf_k4} is an estimate
of the K-band luminosity function inferred from the Sloan Digital Sky
Survey (SDSS) near infrared, 
z$^*$-band luminosity function of Blanton \etal (\shortcite{blanton}).
To convert from z$^*$ (AB system) to standard K we have simply subtracted
2.12~magnitudes from the SDSS z$^*$ magnitudes. This offset consists of 
a contribution of 0.51~magnitudes to 
convert from AB magnitudes to the standard Vega system, and a mean
z$^*$-\K\ colour of 1.61, which we find is typical of the model spectra
discussed in Section~\ref{sec:pop} that match our observed \B-\K\ colours.
As has been noted by Wright (\shortcite{wright}) the luminosity 
function inferred from the SDSS data is offset compared to our 
estimate.
One suggestion put forward by Wright (\shortcite{wright}) is
that the \twomass\ magnitudes could be systematically too faint.
The systematic error would have to amount to 0.5~magnitudes to 
reconcile the luminosity density inferred from the SDSS data with that
which we infer from the \twomass-\twodF\ catalogue. Such an error
is comprehensively excluded by the very small offset 
that was found in Section~\ref{sec:mags} between
the \twomass\ Kron magnitudes and the data of Loveday (\shortcite{love}).
Also, a direct galaxy-by-galaxy comparison of the z$^*$-J and z$^*$-\K\
colours computed using the SDSS Petrosian and \twomass\ Kron magnitudes
produced galaxy colours in good accord with expectations based on model
spectra  (Ivezic, Blanton and Loveday private communication). Finally, we note
that a good match to our estimate of the \K-band luminosity function 
cannot be achieved by simply moving the SDDS curve in 
Fig.~\ref{fig:lf_k4} horizontally. If slid by 0.5~magnitudes to match the 
luminosity density then it falls well below our estimate at bright magnitudes.
However if the SDSS curve is moved vertically, by a factor of 1.6,
then the two estimates come into reasonable agreement at all magnitudes.
Thus, the most likely explanation of the difference between the 
SDSS and \twomass-\twodF\ luminosity functions is the uncertainty in 
the overall normalization induced by large-scale density fluctuations.
It is to be hoped that as the sky coverage of the SDSS and 
\twomass-\twodF\ surveys increases this discrepancy will be reduced.

\subsection{Colour Distributions}
\label{sec:col}

Since our combined \twomass--\twodF\ catalogue includes 
\B-band and infrared magnitudes, it is also possible
 to estimate the \B-band optical luminosity function
and the optical/infrared  bivariate luminosity function.
We do not present the \B-band optical luminosity function here 
as estimates from the \twodF\ are discussed in detail
in Norberg \etal (\shortcite{norberg00}) and decomposed into luminosity
functions of different spectral types in 
Folkes \etal (\shortcite{folkes}) and Madgwick \etal
(\shortcite{madg}). Instead, we present 
the bivariate \B/\K\ and J/\K\ luminosity functions
in Fig.~\ref{fig:col}, in the form of the rest-frame
\B$-$\K\ and J$-$\K\ colour distributions, split by \K-band absolute magnitude.
Results are shown for just our default set of k+e corrections, but the
colour distributions are extremely insensitive to this choice and 
to whether evolutionary corrections are ignored or included.
The shape of \B$-$\K\ colour distribution
varies systematically with \K-band luminosity. At fainter magnitudes there
is an increasingly large population of bluer, star-forming galaxies. The
star formation rate has less effect on the infrared J$-$\K\ colours. 
Here, the shape of the J$-$\K\
colour distribution varies little with luminosity, but the position of the
peak moves gradually redder with increasing luminosity.  Colour
distributions such as these are sensitive to both the distribution of
stellar age and the metallicity, and therefore provide important
constraints on models of galaxy formation (for example, see Cole \etal
\shortcite{cole2000}).

\subsection{Spectral Type Distribution}
\label{sec:sp_type}

\begin{figure}
\centering
\centerline{\epsfysize=8.5 truecm \epsfbox{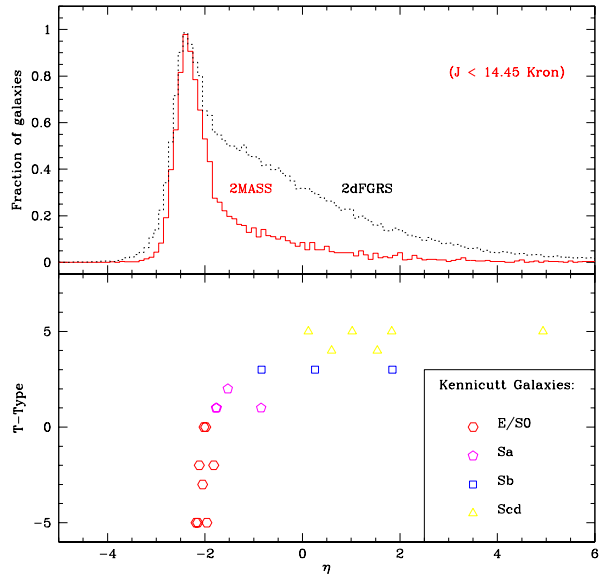}}
\caption{The distribution of the spectral type parameter, $\eta$,
in the full \twodF\ and our matched \twomass\ catalogue (upper panel). 
The lower panel uses a sample of galaxies from Kennicutt (1992) to
show how $\eta$ is correlated with morphological type. 
}
\label{fig:sp_type}
\end{figure}

\begin{figure*}
\centering
\centerline{\epsfbox[0 520 540 750]{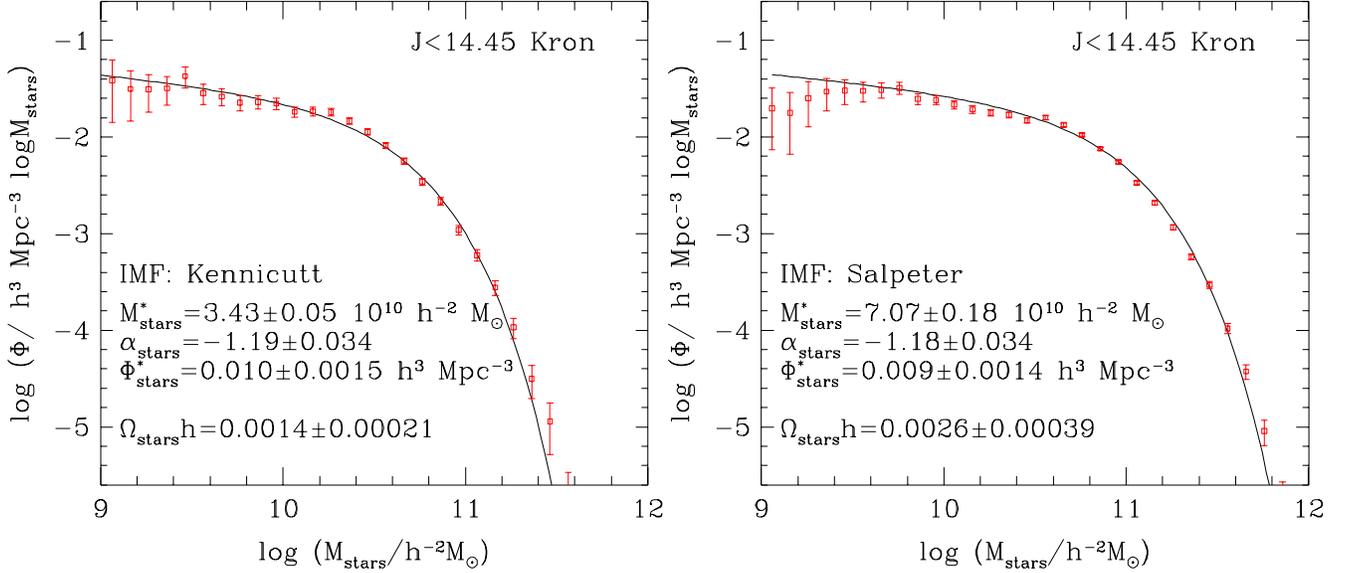}}
\caption{SWML estimates of the stellar mass function (open symbols with
error bars) and
STY Schechter function estimates (lines). The parameter and
error estimates of the Schechter function fits are given in the legends.
The left-hand panel is for a Kennicutt IMF with recycled fraction $R=0.42$  
and the right-hand panel for a Salpeter IMF with $R=0.28$.
}
\label{fig:smf}
\end{figure*}

Another interesting issue that we can address with our data is the 
distribution of spectral types in the \twomass\ catalogue.
For this, we make use of the spectral information in
the 2dF galaxies extracted by a  principal component analysis (Folkes
\etal \shortcite{folkes}).  
Specifically, we use the new continuous variable introduced
in Madgwick, Lahav \& Taylor (\shortcite{mlt}) which is defined by a linear
combination of the first 2 principal component projections, $\eta
\equiv 0.44 pc_1 - pc_2$.  This variable was chosen to be robust to
instrumental uncertainties whilst, at the same time, preserving physical
information about the galaxy.  The dominant influence on the $\eta$
parameter is the relative strength of absorption and emission lines
($\eta$$<$$0$ implies less than average 
emission-line strength while $\eta$$>$$0$ implies stronger than average
emission-line strength).  A more detailed description will be
presented in Madgwick et al. (in preparation).

We can now gain insight into the population mix of our \twomass\
sample by simply creating a histogram of the $\eta$ values for the
corresponding \twomass--\twodF\ matched galaxies with $J$$<$$14.45$
(Kron).  We plot this in Fig.~\ref{fig:sp_type} where we also show data
for the entire \twodF\ sample as comparison.  Also shown in
Fig.~\ref{fig:sp_type} (bottom panel) is the morphology-$\eta$ relation
derived from a sample of galaxy spectra from the Kennicutt Atlas 
(Kennicutt \shortcite{kenn92}).

It can be clearly seen from Fig.~\ref{fig:sp_type} that the
predominant population in the \twomass\ sample is has $\eta$$<$$-2$.  By
contrasting this with values of $\eta$ obtained from the spectra of
galaxies with known morphological type (Kennicutt 1992), we can see
that this corresponds to galaxies of E/S0 morphologies.  More
precisely, the fraction of galaxies in our matched sample with
spectral types corresponding to E/S0 morphologies is 62\% (compared
with $\sim35\%$ in the full \twodF).  Sa-Sb galaxies make up a further
22\% and the remaining 16\% are galaxies of later morphological types.

\subsection{The Galaxy Stellar Mass Function}
\label{sec:smf}

In contrast to optical light, near-infrared luminosities are
relatively insensitive to the presence of young stars and
can be more accurately related to the underlying
stellar mass.  Thus, with relatively few model assumptions, we can derive
the distribution of galaxy stellar masses. The integral of this
distribution is the total mass density in stars, which can be expressed
in units of the critical density as $\Omega_{\rm stars}$.  Attempts to
estimate this quantity date back many decades, but even recent estimates
such as those by  Persic \& Salucci (\shortcite{ps92}), 
Gnedin \& Ostriker (\shortcite{go92}), Fukugita, Hogan \& Peebles 
(\shortcite{fhp98})  and  Salucci \& Persic (\shortcite{sp99})  
have very large uncertainties because they are based on B-band light 
and require uncertain B-band mass-to-light ratios.  
The much more accurate estimate that we provide here should prove
very useful for a variety of purposes.

To estimate the galaxy stellar mass function, we use the modelling of the
stellar populations described in Section~\ref{sec:pop} 
to obtain estimates of the present luminosity and
stellar mass-to-light ratio for each galaxy in the survey.  This
is done on a galaxy-by-galaxy basis as described in Section~\ref{sec:pop}.
The sample we analyze is defined by the $11$$<$J$<$$14.45$ (Kron)
apparent magnitude limits.
The stellar mass that we estimate for each galaxy is the mass locked up in
stars and stellar remnants. This differs from the time integral of the star
formation rate because some of the material that goes into forming massive
stars is returned to the interstellar medium via winds and supernovae. For
a given IMF, this recycled fraction, $R$, can be estimated reasonably
accurately from stellar evolution theory. Here, we adopt the values
$R=0.42$ and $0.28$ for the Kennicutt (\shortcite{kenn83}) and Salpeter
(\shortcite{salpeter}) IMFs respectively, as described in Section~5.2 of
Cole \etal (\shortcite{cole2000}) who made use of the models of Renzini \&
Voli (\shortcite{rv81}) and Woosley \& Weaver (\shortcite{ww95}). Hence,
the stellar masses we choose to estimate are $(1$$-$$R)$ times the time
integral of the star formation rate to the present day.  Note that the IMFs
we consider assume that only stars with mass greater than 0.1M$_\odot$ ever
form and so we are not accounting for any mass that may be 
locked up in the form of brown dwarfs.

\begin{figure}
\centering
\centerline{\epsfxsize=8 truecm \epsfbox[20 190 550 670]{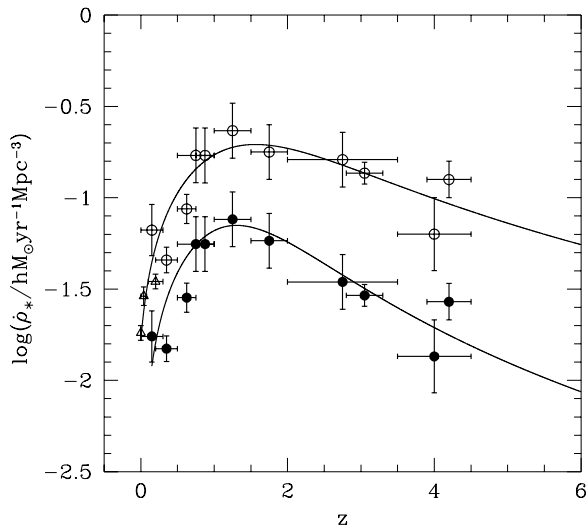}}
\caption{Observational estimates of the star formation history of the
universe. The points with error bars show estimates of the mean star formation
rate per unit volume at various redshifts (see Steidel \etal 1999 and
references therein). The solid symbols are the
star formation rates implied if there is no absorption by dust. The
open symbols show estimates corrected for dust absorption using a
Calzetti (\shortcite{calzetti}) 
extinction law with a mean ${\rm E(B-V)}=0.15$
(Steidel \etal 1999). In both
cases an $\Omega_0=0.3$, $\Lambda_0=0.7$ cosmology has been used
to calculate the volume as a function of redshift and a Salpeter
IMF to convert luminosity to star formation rate.
The smooth curves are the fits we use when integrating over time
to estimate the total mass density of stars formed by the present.
}
\label{fig:madau}
\end{figure}

\begin{table}
\centering
\caption{The SWML stellar mass functions 
as plotted in Fig.~\ref{fig:smf}. 
The units of both $\phi$ and its uncertainty $\Delta \phi$  are number
per $h^{-3}$  Mpc$^3$ per decade of mass.}
\begin{center}
\begin{tabular}{llllll} 
\multicolumn{1}{l} { } &
\multicolumn{1}{l} {Kennicutt}&
\multicolumn{1}{l} {Salpeter}\\ 
\multicolumn{1}{l} {$\log_{10} M$ } &
\multicolumn{1}{l} {$\phi \pm \Delta \phi$} &
\multicolumn{1}{l} {$\phi \pm \Delta \phi$}\\ 
\hline 
 9.06 &   (4.24$\pm$2.62)$\times 10^{-2}$  &   (1.37$\pm$1.05)$\times 10^{-2}$ \\
 9.16 &   (3.42$\pm$1.80)$\times 10^{-2}$  &   (2.41$\pm$1.35)$\times 10^{-2}$ \\
 9.26 &   (3.01$\pm$1.31)$\times 10^{-2}$  &   (2.06$\pm$1.13)$\times 10^{-2}$ \\
 9.36 &   (3.33$\pm$1.11)$\times 10^{-2}$  &   (3.01$\pm$1.13)$\times 10^{-2}$ \\
 9.46 &   (4.21$\pm$1.04)$\times 10^{-2}$  &   (3.25$\pm$0.92)$\times 10^{-2}$ \\
 9.56 &   (2.75$\pm$0.67)$\times 10^{-2}$  &   (2.87$\pm$0.67)$\times 10^{-2}$ \\
 9.66 &   (2.70$\pm$0.55)$\times 10^{-2}$  &   (3.10$\pm$0.56)$\times 10^{-2}$ \\
 9.76 &   (2.31$\pm$0.42)$\times 10^{-2}$  &   (3.30$\pm$0.47)$\times 10^{-2}$ \\
 9.86 &   (2.20$\pm$0.35)$\times 10^{-2}$  &   (2.67$\pm$0.34)$\times 10^{-2}$ \\
 9.96 &   (2.21$\pm$0.31)$\times 10^{-2}$  &   (2.51$\pm$0.27)$\times 10^{-2}$ \\
10.06 &   (1.77$\pm$0.23)$\times 10^{-2}$  &   (2.03$\pm$0.20)$\times 10^{-2}$ \\
10.16 &   (1.91$\pm$0.20)$\times 10^{-2}$  &   (1.93$\pm$0.17)$\times 10^{-2}$ \\
10.26 &   (1.77$\pm$0.16)$\times 10^{-2}$  &   (1.86$\pm$0.15)$\times 10^{-2}$ \\
10.36 &   (1.46$\pm$0.12)$\times 10^{-2}$  &   (1.62$\pm$0.11)$\times 10^{-2}$ \\
10.46 &   (1.11$\pm$0.08)$\times 10^{-2}$  &   (1.49$\pm$0.09)$\times 10^{-2}$ \\
10.56 &   (8.15$\pm$0.61)$\times 10^{-3}$  &   (1.61$\pm$0.08)$\times 10^{-2}$ \\
10.66 &   (5.62$\pm$0.43)$\times 10^{-3}$  &   (1.30$\pm$0.06)$\times 10^{-2}$ \\
10.76 &   (3.39$\pm$0.29)$\times 10^{-3}$  &   (1.06$\pm$0.04)$\times 10^{-2}$ \\
10.86 &   (2.08$\pm$0.20)$\times 10^{-3}$  &   (7.40$\pm$0.30)$\times 10^{-3}$ \\
10.96 &   (1.07$\pm$0.12)$\times 10^{-3}$  &   (5.50$\pm$0.22)$\times 10^{-3}$ \\
11.06 &   (5.95$\pm$0.82)$\times 10^{-4}$  &   (3.29$\pm$0.15)$\times 10^{-3}$ \\
11.16 &   (2.75$\pm$0.49)$\times 10^{-4}$  &   (2.02$\pm$0.10)$\times 10^{-3}$ \\
11.26 &   (1.05$\pm$0.26)$\times 10^{-4}$  &   (1.13$\pm$0.07)$\times 10^{-3}$ \\
11.36 &   (2.77$\pm$1.11)$\times 10^{-5}$  &   (5.56$\pm$0.40)$\times 10^{-4}$ \\
11.46 &   (9.51$\pm$5.65)$\times 10^{-6}$  &   (2.90$\pm$0.26)$\times 10^{-4}$ \\
11.56 &   (2.05$\pm$2.38)$\times 10^{-6}$  &   (9.87$\pm$1.26)$\times 10^{-5}$ \\
11.66 &   (6.87$\pm$13.6)$\times 10^{-7}$  &   (3.73$\pm$0.66)$\times 10^{-5}$ \\
      &                                    &   (8.46$\pm$2.58)$\times 10^{-6}$ \\
      &                                    &   (2.22$\pm$1.20)$\times 10^{-6}$ \\
\hline	      
\end{tabular} 
\end{center}	
\label{tab:smfswml}
\end{table}

Our results are presented in Fig.~\ref{fig:smf} which shows both SWML
and Schechter function estimates of the present-day galaxy stellar
mass function for two choices of IMF. The SWML estimates are tabulated
in Table~\ref{tab:smfswml}. Just as for the luminosity
functions, the stellar mass function is quite well described by the
Schechter functional form.  Integrating over these Schechter functions
to determine the total stellar mass gives $\Omega_{\rm stars}h= (1.4
\pm 0.21) \times 10^{-3}$ for the Kennicutt IMF and $\Omega_{\rm
stars}h= (2.6 \pm 0.39)\times 10^{-3}$ for the Salpeter IMF.  Note
that the integral converges rapidly at both limits and, in particular,
the contribution to $\Omega_{\rm stars}$ from objects with $M_{\rm
stars}<10^9$ h$^{-2}$ M$_\odot$ is negligible.  We find that these
values vary by less than the quoted errors when we alter the assumed
(k+e)-corrections by either ignoring evolution, ignoring dust or
changing $\Omega_0$.  Taken together with our estimates of the \K-band
luminosity density these, estimates imply mean stellar mass-to-light
ratios of $0.73\, \Msun/\Lsun$ in the case of the Kennicutt IMF and
$1.32\, \Msun/\Lsun$ for the Salpeter IMF.  If we apply the correction
we estimated in Section~\ref{sec:mags} to transform \twomass\ Kron
into total magnitudes, then these estimates and their uncertainties
increase to $\Omega_{\rm stars}h= (1.6 \pm 0.24) \times 10^{-3}$ for
the Kennicutt IMF and $\Omega_{\rm stars}h= (2.9 \pm 0.43)\times
10^{-3}$ for the Salpeter IMF.  Both of these estimates are consistent
with the value, 
$\Omega_{\rm stars}= (3.0 \pm 1.0)\times 10^{-3}$ , derived by 
Salucci \&  Persic (\shortcite{sp99}) 
but have fractional statistical errors which are
several times smaller.  With our method, the uncertainty in
$\Omega_{\rm stars}$ is clearly dominated by the uncertainty in the
IMF.  For some purposes, it is not possible to improve upon this
without a more precise knowledge of the true IMF -- assuming there is
a universal IMF.  However, for other applications, such as modelling
the star formation history of the universe, it is necessary to assume
a specific IMF to convert the observational tracers of star formation
to star formation rates.  Hence, in this case, it is the much smaller
statistical errors that are relevant.

\begin{table}
\centering
\caption{Estimates of the present-day mass in stars and stellar remnants
obtained by integrating over observational estimates of the star formation
history of the universe. We express this stellar mass density in terms
of the critical density and give values of $\Omega_{\rm stars} h^2$
estimated for different assumed IMF's and dust corrections. All
values are for an $\Omega_0=0.3$, $\Lambda_0=0.7$ cosmology and assume
stellar populations  of solar metallicity.
}
\begin{center}
\begin{tabular}{lrrrrrrrrrrrrrrrrrr} 
\multicolumn{1}{l} {Dust Extinction} &
\multicolumn{1}{l} {Kennicutt IMF} &
\multicolumn{1}{l} {Salpeter IMF} & \\
\hline 
E(B-V)=0.05 & $0.80 \times 10^{-3}$  &  $1.30 \times 10^{-3}$ \\
E(B-V)=0.10 & $1.17 \times 10^{-3}$  &  $1.86 \times 10^{-3}$ \\
E(B-V)=0.15 & $1.63 \times 10^{-3}$  &  $2.66 \times 10^{-3}$ \\
\hline			        
\end{tabular} 			 
\end{center}
\label{tab:madau}
\end{table}

It is interesting to compare our values with what is inferred by integrating
the observational estimates of the mean star formation history of the
universe. Fig.~\ref{fig:madau} shows observational estimates for one
particular choice of cosmology and IMF and illustrates how the rates
are sensitive to the assumed dust extinction. By fitting a smooth
curve through these estimates, we can calculate the mass of stars formed
by the present day and how this depends on the IMF and
assumed dust extinction. The upper smooth curve shown in Fig.~\ref{fig:madau} 
is of the form $\dot \rho_\star = (a + bz)/(1 + (z/c)^d) h M_\odot $yr$^{-1}$Mpc$^{-3}$, 
where $(a,b,c,d) = (0.0166,0.1848,1.9474,2.6316)$. The data points uncorrected
for dust extinction are fit with $(a,b,c,d) = (0.0,0.0798,1.658,3.105)$.
As for our estimates above, we assume that no mass
goes into forming brown dwarfs and multiply the star formation rate by
$1$$-$$R$, where $R$ is the recycled fraction, so as to form an estimate of
the mass locked up in stars.  Values of $\Omega_{\rm stars} h^2$
estimated in this way are listed in Table~\ref{tab:madau}. The
values in this table are for an $\Omega_0=0.3$, $\Lambda_0=0.7$ cosmology, 
but they are insensitive to this choice. They depend slightly on the 
assumed metallicity of the stellar population and would be 10\%
lower if half solar, rather than solar metallicity were assumed.
Note that the $\Omega_{\rm stars}$ values inferred from the star 
formation history
of the universe scale differently with the assumed Hubble constant than those
inferred above from the IR luminosity functions. For $h=0.7$
our estimates from \twomass\  become
$\Omega_{\rm stars}h^2= (1.12 \pm 0.16)\times 10^{-3} $ for the
Kennicutt IMF and $\Omega_{\rm stars}h^2= (2.03 \pm 0.30)
\times 10^{-3} $ for the Salpeter IMF. Comparison with Table~\ref{tab:madau}
shows that these values are consistent with those  inferred
from the cosmic star formation history only if the dust correction
assumed in the latter is modest, E(B-V)$\approx0.1$. This value is 50\%
smaller than the value preferred by Steidel \etal (\shortcite{steidel99}).

\section{Conclusions}
\label{sec:conc}

The new generation of very large surveys currently underway make it
possible to characterize the galaxy population with unprecedented
accuracy. In this paper, we have combined two such large surveys, the
infrared imaging 2MASS and the 2dF Galaxy Redshift Survey\footnote{A 
table containing the positions, 2MASS infra-red magnitudes
and \twodF\ redshifts used in this paper and electronic versions of
Tables~\ref{tab:lfswml} and~\ref{tab:smfswml} are available at 
http://star-www.dur.ac.uk/$\tilde{\hphantom{n}}$cole/2dFGRS-2MASS .}
to obtain a complete dataset which is more than an 
order of magnitude larger than
previous datasets used for statistical studies of the near-infrared
properties of the local galaxy population.  We have used this combined
catalogue to derive the most precise estimates to date of the galaxy J and
\K-band luminosity functions and of the galaxy stellar mass function.

Characterizing the near-infrared properties of galaxies offers several
advantages. Firstly, the near-infrared light is dominated by established,
old stellar populations rather than by the recent star formation activity
that dominates the blue light. Thus, the J and K-band luminosity functions
reflect the integrated star formation history of a galaxy and, as a result,
provide particularly important diagnostics of the processes of galaxy
formation. For the same reason, the distribution of stellar mass in
galaxies --the galaxy stellar mass function-- can be derived from the
near-infrared luminosities in a relatively straightforward way, with only a
weak model dependence. Finally, corrections for dust extinction as well as
k-corrections are much smaller in the near-infrared than in the optical. 

Due to the size of our sample, our determination of the J- and
\K-band galaxy luminosity functions have, for the most part, 
smaller statistical errors than previous estimates. Furthermore, since our
sample is infrared-selected, our estimates are free from any potential
biases that might affect infrared luminosity functions derived from
optically-selected samples. We find that the J- and \K-band galaxy
luminosity functions are fairly well described by Schechter functions,
although there is some evidence for an excess of bright galaxies
relative to the best-fit Schechter functional form. In general, the
SWML estimates are a truer representation of the luminosity functions. 
Our K-band estimates are in overall agreement
with most previous determinations, but have smaller statistical errors.

The exception is the K-band luminosity function inferred from the 
near infrared SDDS photometry (\cite{blanton}). The difference between
the K-band luminosity function we infer from their data and our own
estimate is too large to be explained by photometric differences.  The
difference between the two estimates is better described by a
difference in overall number density of a factor of $1.6$.  A similar
discrepancy is seen in the \B-band between the SDSS and \twodF\
luminosity function estimates (see \cite{norberg00} and
\cite{blanton}).  The suspicion is that the uncertainty in the overall
normalization of the luminosity functions induced by large-scale
structure within the large, but finite, survey volumes could be to
blame. However, the errors that we quote for the \twodF-\twomass\
luminosity functions already include an estimate of this sampling
uncertainty as derived from realistic mock catalogues.  A similar
exercise for a catalogue with the same area and depth as that of
Blanton \etal (\shortcite{blanton}) indicates that the required
overdensity of a factor of $1.6$ is unlikely.  So probably there is
more than one contributory factor at work and the hope is that these
will be identified as the surveys progress.

Using our J-band luminosity function, \B$-$\K\ and J$-$\K\ colours and
simple galaxy evolutionary tracks, we have obtained the first estimate
of the galactic stellar mass function derived directly from
near-infrared data. We find that this mass function is also fairly
well described by a Schechter form. An integral over the stellar mass
function gives $\Omega_{\rm stars}$, the universal mass density locked
up in luminous stars and stellar remnants, expressed in terms of the
critical density.  $\Omega_{\rm stars}$ is a key component of the
overall inventory of baryons in the universe. An accurate
determination of this quantity is essential for detailed comparisons
with other quantities of cosmological interest such as the total
baryonic mass, $\Omega_{\rm baryon}$, inferred from Big Bang
nucleosynthesis considerations (e.g. Burles \& Tytler 1998), the
cosmic star formation rate (e.g. Steidel \etal \shortcite{steidel99}),
and the cosmic evolution of the gas content of the universe
(Storrie-Lombardi \& Wolfe 2000).  The statistical uncertainty in our
estimate of $\Omega_{\rm stars}$ is about 15\%, several times 
smaller than the best previous determination by Salucci \& Persic 
(\shortcite{sp99}).
In fact, the errors in $\Omega_{\rm stars}$ are
dominated by systematic uncertainties associated with the choice of
stellar IMF in the galaxy evolution model. A Kennicutt IMF gives
$\Omega_{\rm stars}h= (1.6 \pm 0.24)\times 10^{-3}$ while a Salpeter
IMF gives $\Omega_{\rm stars}h= (2.9 \pm 0.43)\times 10^{-3}$.  For
$h=0.7$, these values correspond to less than 11\% of the baryonic
mass inferred from Big Bang Nucleosynthesis (\cite{burles}).  Our
values of $\Omega_{\rm stars}$ today are only consistent with recent
determinations of the integrated cosmic star formation if the
correction for dust extinction is modest.

\section*{Acknowledgements}
We thank John Lucey for the initial suggestion to look at the
\twomass\ database. We thank Tom Jarrett for advice and detailed
explanations of the 2MASS data reduction procedure. We also thank Ian
Smail for help in manipulating DSS images, Jon Loveday for supplying
his luminosity function data in electronic form and Simon White for
useful suggestions. We are grateful to the VIRGO consortium for
allowing us to use the Hubble Volume simulation data prior to
publication.The redshift data used here were obtained with the
2-degree field facility on the 3.9m Anglo\-Australian Telescope
(AAT). We thank all those involved in the smooth running and continued
success of the 2dF and the AAT. SMC acknowledges a PPARC Advanced
Fellowship, IPRN an SNSF and ORS Studentship and CSF a Leverhulme
Research Fellowship.

The 2dF Galaxy Redshift Survey was made possible through the dedicated
efforts of the staff at the Anglo-Australian Observatory, both in creating
the two-degree field facility and supporting it on the telescope.

\end{document}